\documentclass[12pt, appendixfloats, numberedappendix]{emulateapj}

\def\eg{e.g., }

\newcommand{\Mpc}{{\rm\,Mpc}}
\newcommand{\mpc}{{\rm\,Mpc}}
\newcommand{\kpc}{{\rm\,kpc}}

\newcommand{\beq}{\begin{equation}}
\newcommand{\eeq}{\end{equation}}

\def\mgi{\ion{Mg}{1}}
\def\mgii{\ion{Mg}{2}}

\def\mgiidoublet{\ion{Mg}{2}~$\lambda\lambda2796,2803$}

\def\alii{\ion{Al}{2}}

\def\cii{\ion{C}{2}}

\def\civ{\ion{C}{4}}

\def\siii{\ion{Si}{2}}

\def\siiv{\ion{Si}{4}}

\def\caii{\ion{Ca}{2}}

\def\lymana{Ly$\alpha$}

\newcommand{\kms}{\ensuremath{{\rm km~s}^{-1}}}

\newcommand{\rewmgiitwo}{\ensuremath{W_0^{\lambda2803}}}
\newcommand{\rewmgii}{\ensuremath{W_0^{\rm Mg\,II}}}

\newcommand{\MSun}{\ensuremath{\rm M_\odot}}

\newcommand{\mgplus}{\ensuremath{\rm{Mg^+}}}

\newcommand{\rvir}{\ensuremath{r_{\rm vir}}}
\newcommand{\ud}{\ensuremath{{\rm d}}}

\usepackage{natbib}
\usepackage{rotating}
\usepackage{amsmath}

\begin{document}

\title{The Large-scale Distribution of Cool Gas around Luminous Red Galaxies}
\shorttitle{The large-scale distribution of \mgii}
\shortauthors{Zhu, M\'enard et al.}

\author{
Guangtun Zhu\altaffilmark{1}, Brice M{\'e}nard\altaffilmark{1,2,3}, \\
Dmitry Bizyaev\altaffilmark{4},
Howard Brewington\altaffilmark{4},
Garrett Ebelke\altaffilmark{4},
Shirley Ho\altaffilmark{5},\\
Karen Kinemuchi\altaffilmark{4},
Viktor Malanushenko\altaffilmark{4},
Elena Malanushenko\altaffilmark{4},
Moses Marchante\altaffilmark{4},\\
Surhud More\altaffilmark{2},
Daniel Oravetz\altaffilmark{4},
Kaike Pan\altaffilmark{4},
Patrick Petitjean\altaffilmark{6},
Audrey Simmons\altaffilmark{4}
} 
\altaffiltext{1}{Department of Physics \& Astronomy, Johns Hopkins University, 3400 N. Charles Street, Baltimore, MD 21218, USA, gz323@pha.jhu.edu}
\altaffiltext{2}{Kavli IPMU (WPI), the University of Tokyo, Kashiwa 277-8583, Japan}
\altaffiltext{3}{Alfred P. Sloan Fellow}
\altaffiltext{4}{Apache Point Observatory and New Mexico State University, P.O. Box 59, Sunspot, NM, 88349-0059, USA}
\altaffiltext{5}{Bruce and Astrid McWilliams Center for Cosmology, Department of Physics, Carnegie Mellon University, 5000 Forbes Ave, Pittsburgh, PA 15213, USA}
\altaffiltext{6}{Institut d'Astrophysique de Paris, UPMC-CNRS, UMR7095, 98bis Boulevard Arago, 75014 - Paris, France}

\begin{abstract}
We present a measurement of the correlation function between luminous red galaxies and cool gas traced by \mgiidoublet\ absorption, on scales ranging from about 30 kpc to 20 Mpc. The measurement is based on cross-correlating the positions of about one million red galaxies at $z\sim0.5$ and the flux decrements induced in the spectra of about $10^5$ background quasars from the Sloan Digital Sky Survey.  We find that:
(i) This galaxy-gas correlation reveals a change of slope on scales of about 1 Mpc, consistent with the expected transition from a dark matter halo dominated environment to a regime where clustering is dominated by halo-halo correlations.
Assuming that, on average, the distribution of \mgii\ gas follows that of dark matter up to a gas-to-mass ratio, we 
find the standard halo model to provide an accurate description of the gas distribution over three orders of magnitude in scale. Within this framework we estimate the average host halo mass of luminous red galaxies to be about $10^{13.5}\,\MSun$, in agreement with other methods. We also find the \mgii\ gas-to-mass ratio around LRGs to be consistent with the cosmic value estimated on Mpc scales. 
Combining our galaxy-gas correlation and the galaxy-mass correlation function from galaxy-galaxy lensing analyses we can directly measure the \mgii\ gas-to-mass ratio as a function of scale and reach the same conclusion.
(ii) From line-width estimates, we show that the velocity dispersion of the gas clouds also shows the expected 1- and 2-halo behaviors. 
On large scales the gas distribution follows the Hubble flow, whereas on small scales we observe the velocity dispersion of the \mgii\ gas clouds to be lower than that of collisionless dark matter particles within their host halo. This is in line with the fact that cool clouds are subject to the pressure of the virialized hot gas.
This work highlights the potential of galaxy-gas correlations as a powerful tool to probe the cosmic baryon cycle and the large-scale distribution of metals. 
\end{abstract}


\section {Introduction}

Understanding the large-scale distribution of matter is a major goal in astrophysics. The advent of large photometric sky surveys combined with statistical analyses has allowed us to characterize the distribution of stars, dark matter and dust well beyond galactic disks. However, the large-scale distribution of gas and in particular gaseous metals which encodes key information about the cosmic baryon cycle remains poorly constrained.

Absorption line spectroscopy has been used for more than three decades to probe the distribution of gas around galaxies, the circumgalactic medium (CGM). Analyses have typically focused on the study of individual absorbers detected in the spectra of background quasars. While this approach has its merit, it is restricted to the study of strong absorbers and only allows us to probe the tip of the iceberg of the overall gas distribution. Probing the matter distribution on large scales where density is low requires a large range in sensitivity, which statistical analyses can often offer. Such statistical approaches have been succesfully applied numerous times to broad-band photometric surveys. However, statistical analyses aimed at probing the gaseous content of the CGM with spectroscopic data by extracting information below the noise level of individual spectra have been limited to a handful of analyses \citep{steidel10a, bordoloi11a, zhu13b} constraining the gas distribution within a few hundred kpc around galaxies.

When measured over a broad range of scales, spatial correlation functions can provide us with valuable information on the distribution of matter within and beyond dark matter halos. Obtaining such a measurement in the context of galaxy-gas correlations requires (i) a large number of foreground galaxies and background sources and (ii) the presence of an abundant species giving rise to a strong absorption feature. With existing datasets, maximizing those two criteria can be done by selecting LRGs from the Sloan Digital Sky survey \citep[SDSS,][]{york00a, eisenstein11a} as foreground objects and measuring the associated \mgii\ absorption. 
In this paper we present results of an analysis aimed at 
using these samples to measure the galaxy-gas correlation function over a broad range of  scales.
The measurement is based on a spatial cross-correlation between the position of about one million luminous red galaxies (LRGs) at $z\sim0.5$ from SDSS and flux fluctuations induced in the spectra of background quasars by \mgii\ absorption lines. This measurement allows us to characterize the gaseous density profile on scales ranging from the inner dark matter halo of the galaxies up to more than ten megaparsecs where the Hubble flow dominates the dynamics of galaxies\footnote{In an independent analysis, P\'erez-R\`afols et al. (in prep) also detect the galaxy-MgII absorption correlation up to Mpc scales with similar amplitude. The authors use this signal to estimate the cosmic opacity due to \mgii\ absorption.}.

The paper proceeds as follows: we introduce the formalism of galaxy-gas correlation function in Section~\ref{sec:method} and the datasets in Section~\ref{sec:dataanalysis}. The measurements are presented in Section~\ref{sec:results} and we discuss the results in the context of standard cold dark matter (CDM) paradigm in Section~\ref{sec:halomodel}. Section~\ref{sec:summary} summarizes our findings. 
Throughout this work we assume the $\Lambda$CDM cosmology with $(\Omega_{\rm m}, \Omega_{\rm \Lambda}, h, \sigma_8, n_{\rm s})=(0.3, 0.7, 0.7, 0.8, 0.96)$. The Roman subscript `$\mathrm{m}$' stands for all matter and unless stated otherwise scales are in physical units.

\section{Formalism}\label{sec:method}


The spherically-averaged galaxy-gas spatial correlation function is defined as
\beq
\xi_{\rm gal-gas} (r_{\rm 3D}) \equiv \langle \delta_{\rm gal} (r'_{\rm 3D}) \cdot \delta_{\rm gas} (r'_{\rm 3D}+r_{\rm 3D}) \rangle\,\mathrm{,}
\eeq
where $\delta$ is the density contrast, $\delta \equiv \rho/\overline{\rho}-1$, and the ensemble average is performed over the entire survey volume. 
The projected correlation function is given by
\beq
\omega_{\rm gal-gas} (r_{\rm p}) \equiv \left \langle \delta_{\rm gal} (r') \cdot \delta_{\rm gas} (r'+r_{\rm p}) \right \rangle \,\mathrm{,}
\label{eq:projectedomega}
\eeq
where the 2-dimensional density contrast is defined as $\delta \equiv \Sigma/\overline{\Sigma}-1$ and the surface density $\Sigma$ is the integral of 3D density $\rho$ along the line of sight over a redshift path of interest, and the ensemble average is performed over the entire survey area. When the galaxy field is discretized, i.e. when one considers only the positions of galaxy centers, the galaxy density contrast is given by a series of Dirac functions $\delta_D(r'-r'_i)$ at the position of each galaxy $i$. This restricts the ensemble average of the above equation to the positions of galaxies. The cross-correlation then reads
\beq
\omega_{\rm gal-gas} (r_{\rm p}) = \left \langle \frac{\Sigma_{\rm gas}^{\rm tot}  (r_{\rm p}) - \overline{\Sigma}_{\rm gas}}{\overline{\Sigma}_{\rm gas}} \right \rangle_{\rm gal}\;.
\eeq
The \emph{total} mean gas surface density around galaxies can be expressed as
\beq
\langle \Sigma_{\rm gas}^{\rm tot}  (r_{\rm p}) \rangle_{\rm gal} = \overline{\Sigma}_{\rm gas}\,\left[\omega_{\rm gal-gas} (r_{\rm p}) +1 \right]\;.
\eeq
In this work we constrain the galaxy-gas correlation by measuring the \emph{relative} gas absorption along quasar sightlines probing the vicinity of galaxies with respect to reference quasars. We are therefore not sensitive to the background value of the gas surface density and our analysis only allows us to measure the \emph{excess} gas surface density around galaxies, $\Sigma_{\rm gas}$. This is given by
\beq
\langle \Sigma_{\rm gas} (r_{\rm p}) \rangle_{\rm gal} \equiv \overline{\Sigma}_{\rm gas}\,\omega_{\rm gal-gas} (r_{\rm p})\;.
\label{eq:surfacedensity}
\eeq
The projected surface gas density of a given species~$X$ is given by the product of its atomic mass $m_X$ and column denstiy $N$
\beq
\Sigma_X = N \times m_X \,\mathrm{.}
\eeq
The absorption by atoms in the gas phase induces an optical depth $\tau(\lambda)$ given by
\beq
\tau(\lambda) = \frac{\pi e^2}{m_e c}fN\phi \left[\nu(\lambda)\right] \,\mathrm{,}
\eeq
which is proportional to the column density $N$, oscillator strength $f$, and line profile $\phi(\nu)$. For a single-cloud system, the line profile follows the Voigt form determined by the transition wavelength $\lambda_0$, the intrinsic Lorentz width $\gamma$, the Doppler broadening factor $b$ and the line-of-sight velocity $V_{0}$. For a single-cloud system, the center-of-line optical depth is approximately
\beq
\tau_0 \simeq 1.5\times10^{-2} 
\, \frac{Nf\lambda}{b} \,\mathrm{,}
\eeq
where $N$ is in unit of $\mathrm{cm}^{-2}$, $\lambda$ in \AA, and $b$ in $\kms$.
For a multi-cloud system, the line profile also depends on the number of clouds and their velocity spread. The optical depth causes a flux decrement in the background source spectrum given by
\beq
R(\lambda) \equiv \frac{F(\lambda)}{\hat F_{\rm cont}(\lambda)} = e^{-\tau(\lambda)} \, \mathrm{,}
\label{eq:residual}
\eeq
where $F(\lambda)$ is the observed spectrum and $\hat F_{\rm cont}(\lambda)$ is the intrinsic continuum of the background source. From an observational point of view, we quantify the optical depth by measuring the absorption rest equivalent width $W_0$, obtained by integrating the flux decrement over the absorption line profile defined by $\phi(\lambda)$,
\begin{eqnarray}
W_0 & \equiv & \int[1-e^{-\tau(\lambda)}]\,\ud \lambda \nonumber \\
 & = & \int[1-R(\lambda)]\,\ud \lambda\,\mathrm{.}
\end{eqnarray}
If the optical depth at the line center is smaller than unity, the column density is simply given by 
\beq
N = 1.13 \times 10^{20}\,\mathrm{cm}^{-2}\,\frac{W_0}{f\lambda^2}\, \mathrm{,}
\label{eq:curveofgrowth}
\eeq
where both $W_0$ and $\lambda$ are in unit of \AA.

The above equations show that the projected galaxy-gas correlation function can be constrained by measuring the correlation between galaxy positions and the rest equivalent width induced by its surrounding gas distribution
\begin{eqnarray}
\left \langle W_0 \right \rangle_{\rm gal}  (r_\mathrm{p}) & \equiv & \left \langle \delta_\mathrm{gal} (r) \cdot W_0 (r+r_\mathrm{p}) \right \rangle \nonumber \\
 & = & \int[1-\left \langle R(\lambda,r_p) \right \rangle_{\rm gal}]\,\ud \lambda\,\mathrm{.}
\label{eq:meanrew}
\end{eqnarray}
In the next sections we will present a measurement of $\left \langle W_0 \right \rangle_{\rm gal}  (r_\mathrm{p})$ for \mgii\ absorption induced by gas around LRGs. 
In the rest of the paper all scale-dependent ensemble averages will be taken around galaxies. For clarity we will drop the subscript 'gal' in the formalism.

\begin{figure*}
\epsscale{0.8}
\plotone{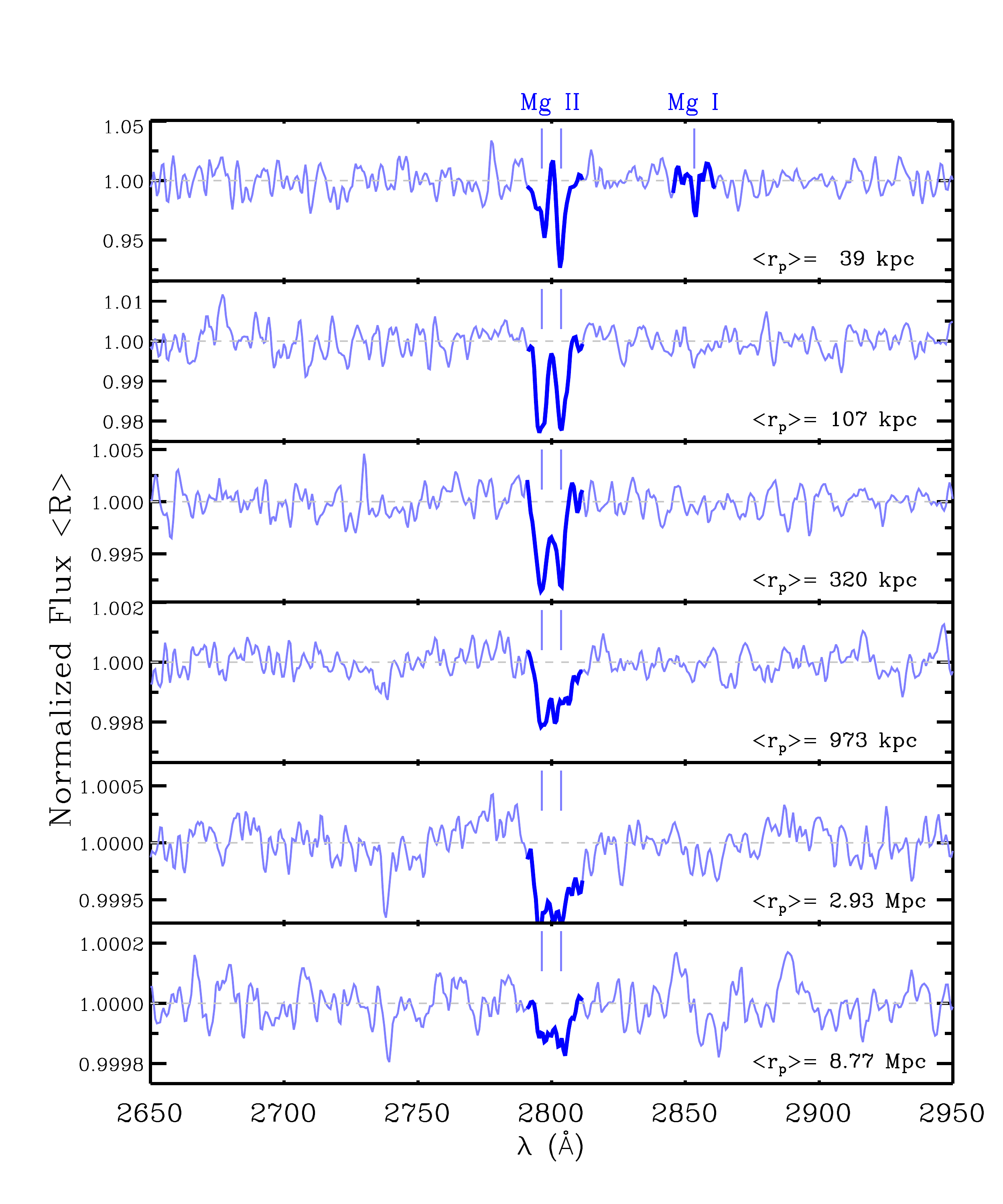}
\caption{Examples of stacked continuum-normalized spectra of background quasars as a function of impact parameter (projected galactocentric distance) from foreground luminous red galaxies (LRGs) at $z\sim0.5$. The vertical ticks and dark blue colors mark the expected positions of \mgiidoublet\ and \mgi$\,\lambda2853$.
}
\label{fig:stackspec}
\end{figure*}

\begin{figure*}[t]
\epsscale{1.2}
\plotone{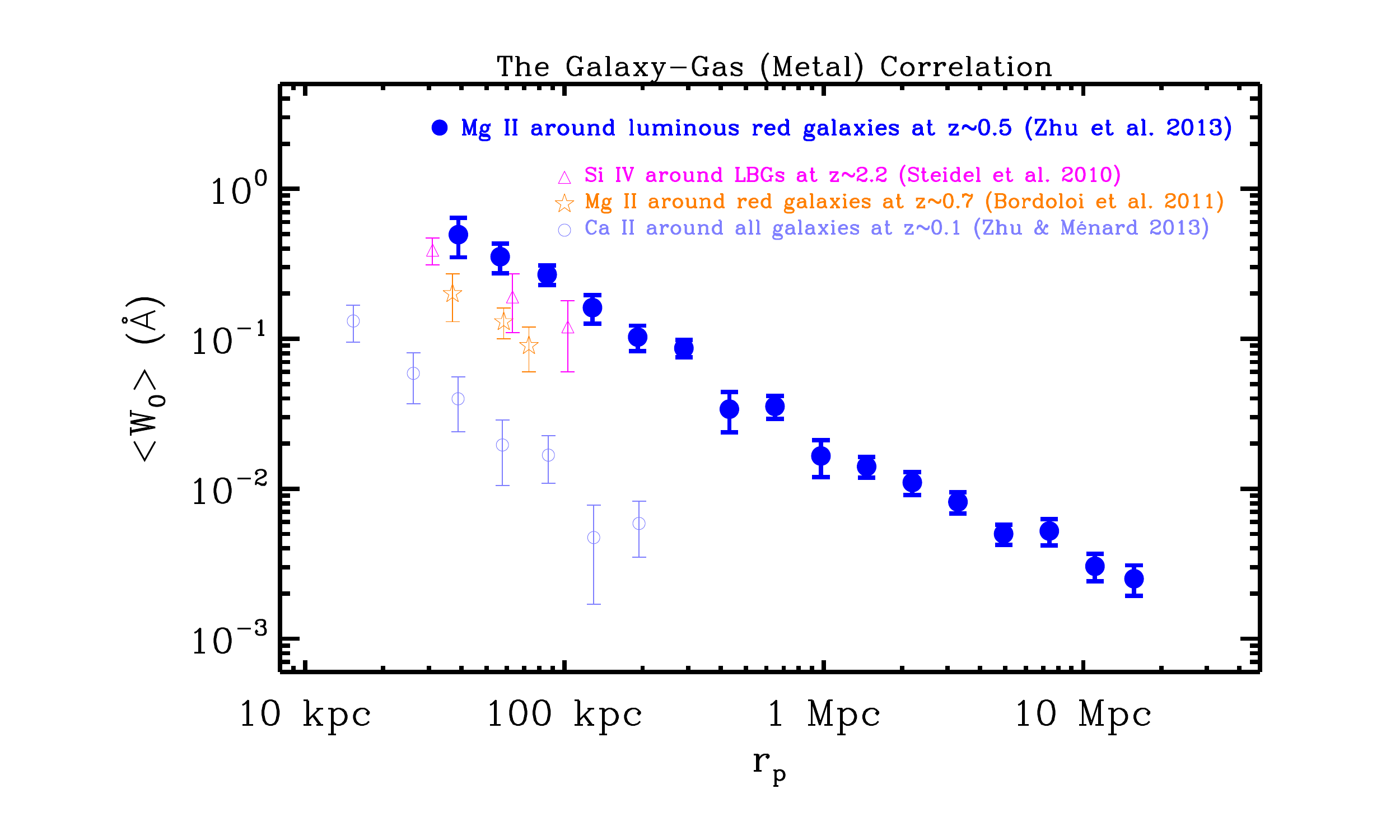}
\hspace{-1.0in}
\vspace{-0.5cm}
\caption{Mean gas absorption profiles (in terms of rest equivalent widths) as a function of scale. The blue solid circles represent our measurements of the LRG-\mgii\  correlation function at $z\sim0.5$ (here the quoted rest equivalent width corresponds to the sum of the two lines $\lambda\lambda2796,2803$). It is detected from about 30 kpc to 20 Mpc. Other symbols show measurements of several metal species around different types of galaxies from the literature (see the text).}
\label{fig:correlation}
\end{figure*}

\section{Data analysis}\label{sec:dataanalysis}

Our goal is to constrain the galaxy-gas (metal) correlation function over a broad range of scales. Doing so requires (i) a large number of foreground galaxies and background sources and (ii) the presence of an abundant species giving rise to a strong absorption feature. With existing datasets, maximizing those two criteria is done by selecting LRGs from the SDSS as foreground objects and measuring the associated \mgii\ absorption. 

\subsection{\mgii\ absorption lines}\label{sec:mgii}

The \mgiidoublet\ doublet has played a major role in gas astrophysics because of their strength and their location in the visible part of the spectrum. They correspond to the fine structure splitting of the singly ionized magnesium excited states \mgii\ (\mgplus). Being an abundant element, $\log(\rm{Mg/H})_\odot+12 \simeq 7.6$ \citep{asplund09a}, it is found in a range of astrophysical environments. Magnesium is a moderately refractory element and has ionization potentials of $7.65$ and $15.04\,$eV, for \mgi\ and \mgii, respectively \citep{morton03a}. At redshift greater than about 0.3, the \mgiidoublet\ lines are the strongest absorption lines of $10^4$ K gas accessible to ground-based observations. The \mgii\ doublet has been used for three decades to study the intergalactic medium. It is the lines used in the observational discovery of the CGM \citep[][]{bergeron86a} and has been used extensively since then \citep[\eg][among others]{steidel92a, churchill99a, nestor05a, narayanan07a}.

The oscillator strength of the two lines are $0.608$ and $0.303$ for \mgiidoublet\ (Kelleher \& Podobedova 2008). When both lines are saturated, their line ratio is one, and when neither is saturated, the line ratio is two. 
For a thermal broadening factor $b$ is about $4\,\kms$ \citep[corresponding to about $25,000\,\mathrm{K}$, \eg][]{churchill00a}, saturation begins for a \mgii\ column density of about $10^{12.5}\,{\rm cm}^{-2}$ which occurs at a total rest equivalent width ($\rewmgii$, sum of the two lines) of about $ 0.15\,$\AA.

\subsection{Samples and analysis}

\begin{figure*}
\hspace{-0.3cm}
\epsscale{0.58}
\plotone{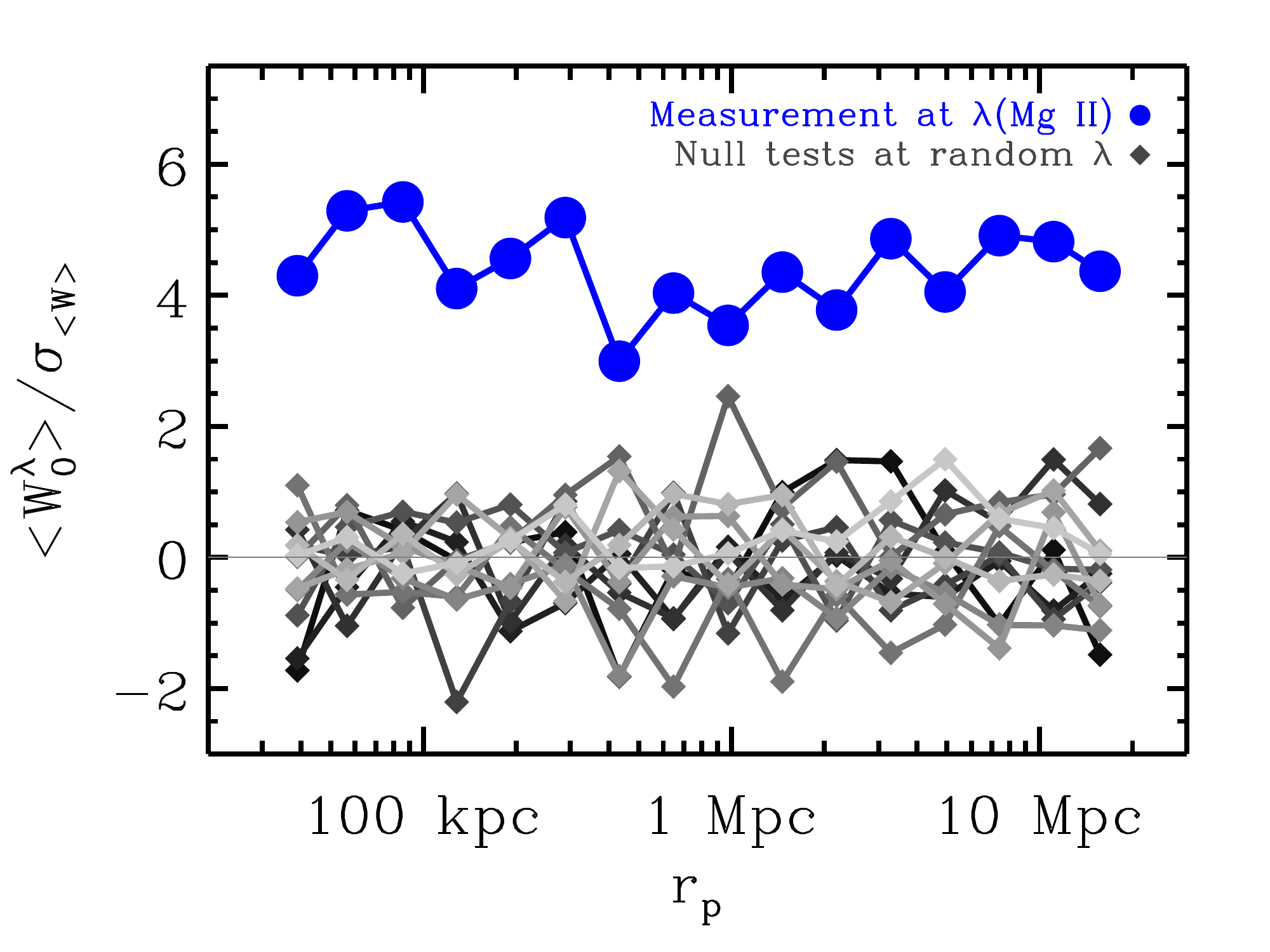}
\hspace{-0.3cm}
\epsscale{0.58}
\plotone{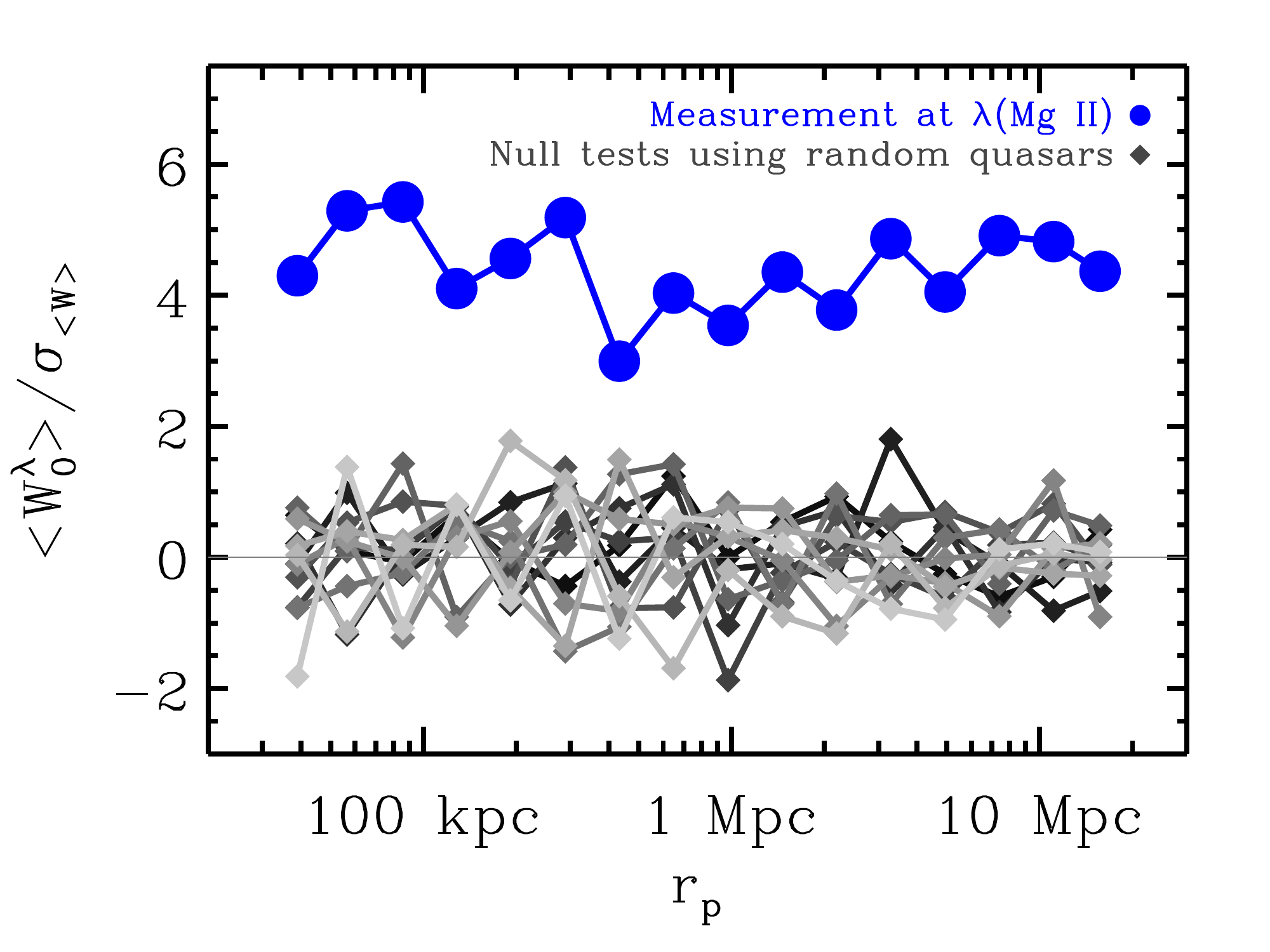}
\caption{Null hypothesis tests for the robustness of the \mgii\ detection. {\it Left} panel: significance of rest equivalent width measurements at randomly-selected wavelengths. {\it Right} panel: significance of rest equivalent width measurements using random quasars at the same redshifts as those in galaxy-quasar pairs. The blue solid circles show the significance of the \mgii\ absorption measurements.}
\label{fig:nulltest}
\end{figure*}

The sample of LRGs used in this work originates from the eleventh Data Release (DR11\footnote{DR11 will be released in December 2014. Here we use the redshift catalog based on version v$5\_6\_0$ of the reduction pipeline \citep{bolton12a}.}) of SDSS. It includes about one million LRGs from the Baryonic Oscillation Spectroscopic Survey \citep[BOSS,][]{dawson13a} with mean stellar mass $\langle M_* \rangle \sim10^{11.5}\,\MSun$ \citep[\eg][]{chenYM12a} and redshift $\langle z \rangle \sim0.57$. The photometric and spectroscopic data were obtained with the wide-field imaging camera \citep{gunn98a} and the new multi-object spectrographs \citep{smee13a} on the SDSS telescope \citep{gunn06a}. From this parent sample we select $849,534$ galaxies at $0.4<z<0.75$ where \mgii\ is accessible in the optical. We do not have additional selection requirement other than the redshift cut and therefore almost all the galaxies are optically luminous and red.

We measure the absorption induced by the gas around these galaxies in the spectra of background quasars. We use spectra from the Data Release 7 \citep[DR7,][]{abazajian09a, schneider10a} of SDSS I-II and the improved redshift estimates by \citet{hewett10a}. The sample includes $107,194$ quasars at $0.1<z<6.5$. 

Accurate estimation of the source flux continuum $\hat F(\lambda)$ (Equation~\ref{eq:residual}) is crucial to detect absorption features. We use the method presented in \citet[][]{zhu13a}, applied to the $84,533$ quasars with $z<4.7$. In a nutshell, this method employs the robust dimensionality-reduction technique {\it nonnegative matrix factorization} \citep[NMF,][]{lee99a, blanton07a} to construct a basis set of nonnegative quasar eigenspectra, and fits each observed quasar spectrum with a nonnegative linear combination of these eigenspectra. Large-scale residuals not accounted for by the NMF basis set are removed with appropriate median filters. The smallest width of such filters has to be kept significantly larger than the size of the absorption feature we are interested in. 
This set of flux residuals has been used to create a sample of about 50,000 absorber systems \citep{zhu13a} and to measure the total amount of \caii\ around low-redshift galaxies \citep{zhu13b}. In the present analysis we use only quasars for which $z_{\rm quasar}-z_{\rm LRG}>0.1$. The median stellar mass and redshift of LRGs in the LRG-quasar pairs are $\langle M_* \rangle = 10^{11.4}\,\MSun$ and $\langle z \rangle \simeq0.52$. 
The set of flux residuals obtained this way allows us to construct composite residuals consistent with unity at the one percent level. To further improve the accuracy and remove systematic trends, we apply our procedure to a set of LRG-quasar pairs for which the quasars are selected to have the same redshift distribution as the original sample but are randomly selected over the sky. This is used to map out large-scale, sub-percent systematic shifts in the mean residuals which are then subtracted when analyzing a given sample. This step is required to properly estimate the zero point of the mean flux residuals over a broad wavelength range.


To quantify the rest equivalent width of the absorption of the \mgiidoublet\ doublet, we perform a double-Gaussian fit of the absorption feature expected at the redshift of the galaxy, allowing the width and line ratio to be free parameters. Absorption being a multiplicative effect we estimate the ensemble average using a geometric mean. This provides us with an estimate of the arithmetic mean of the corresponding optical depth. However, we note that using an arithmetic mean yields similar results, as expected when measuring weak absorption lines. Our estimator is inverse-variance weighted, using the wavelength-dependent noise given by the SDSS pipeline. Throughout the paper, \emph{we will present the total rest equivalent width of the doublet instead of just one of the two lines}.

\section{Results}\label{sec:results}

\subsection{The galaxy-gas correlation}\label{sec:correlation}

\begin{deluxetable}{ccccc}
\tabletypesize{\scriptsize}
\tablecolumns{5}
\tablecaption{The LRG-Mg II correlation at $z\sim0.5$}
\tablehead{
 \colhead{$r_\mathrm{p}$ bin} &  \colhead{Median $r_\mathrm{p}$} &  \colhead{$N_{\rm pairs}$} & \colhead{$\langle W_0^{\rm Mg\,II} \rangle$\tablenotemark{$a$}} & \colhead{$\sigma(\langle W_0^{\rm Mg\,II} \rangle)$\tablenotemark{$b$}} \\ 
 \colhead{[Mpc]} &  \colhead{[Mpc]} & \colhead{ } & \colhead{[m\AA]} & \colhead{[m\AA]} \\ 
}
\startdata
($ 0.030, 0.045$]\tablenotemark{$c$} & $ 0.039$ & $     35$ & $ 494.71 $ & $   145.21$ \\
($ 0.045, 0.068$] & $ 0.056$ & $     88$ & $ 352.58 $ & $    78.68$ \\
($ 0.067, 0.101$] & $ 0.086$ & $    200$ & $ 267.94 $ & $    40.05$ \\
($ 0.101, 0.152$] & $ 0.128$ & $    434$ & $ 161.29 $ & $    34.91$ \\
($ 0.152, 0.228$] & $ 0.191$ & $    880$ & $ 102.67 $ & $    19.76$ \\
($ 0.228, 0.342$] & $ 0.289$ & $   1936$ & $  86.60 $ & $    11.49$ \\
($ 0.342, 0.513$] & $ 0.432$ & $   3964$ & $  33.95 $ & $    10.10$ \\
($ 0.513, 0.769$] & $ 0.648$ & $   8911$ & $  35.42 $ & $     6.11$ \\
($ 0.769, 1.153$] & $ 0.974$ & $  19981$ & $  16.54 $ & $     4.59$ \\
($ 1.153, 1.730$] & $ 1.461$ & $  45030$ & $  14.06 $ & $     2.20$ \\
($ 1.730, 2.595$] & $ 2.192$ & $ 101153$ & $  11.01 $ & $     1.94$ \\
($ 2.595, 3.892$] & $ 3.287$ & $ 228261$ & $   8.17 $ & $     1.32$ \\
($ 3.892, 5.839$] & $ 4.929$ & $ 512263$ & $   5.00 $ & $     0.77$ \\
($ 5.839, 8.758$] & $ 7.395$ & $1151523$ & $   5.23 $ & $     1.03$ \\
($ 8.758,13.137$] & $11.092$ & $2591671$ & $   3.04 $ & $     0.63$ \\
($13.137,18.000$] & $15.694$ & $4086471$ & $   2.51 $ & $     0.57$ \\

\enddata
\tablenotetext{a}{Mean rest equivalent width of \mgii\ (sum of two lines).}
\tablenotetext{b}{Bootstrapping errors of $\langle W_0^{\rm Mg\,II} \rangle$.}
\tablenotetext{c}{Mg I measurement in this bin: $\langle W_0^{\rm Mg\,I} \rangle=83\pm64\,$m\AA.}
\label{tbl:correlation}
\end{deluxetable}


We measure the spatial cross-correlation between the position of our selected sample of LRGs and the \mgii\ rest equivalent width induced in the spectra of background quasars, as a function of scale, $\langle \rewmgii \rangle (r_\mathrm{p})$ (see Eq~\ref{eq:meanrew}). 
Figure~\ref{fig:stackspec} presents examples of the intermediate products of the analysis, the stacked continuum-normalized spectra $\langle R(\lambda) \rangle$. The figure highlights the expected positions of \mgiidoublet\ and \mgi$\,\lambda2803$ with vertical tick marks and dark blue color. Note that the absorption scale varies from about $10^{-2}$ at the top to about $10^{-4}$ at the bottom. In Table~\ref{tbl:correlation} and Figure~\ref{fig:correlation}, we present the mean \mgii\ rest equivalent width $\langle W_0 \rangle$ (including the contribution from both absorption lines) with solid circles between $30\,\kpc$ and $20\,\Mpc$.
We estimate the rest equivalent width errors by bootstrapping the sample of LRG-quasar pairs one hundred times.

To validate the robustness of these measurements, we perform two null hypothesis tests: (1) we measure the mean rest equivalent width at randomly chosen wavelengths; and (2) we measure the expected \mgii\ rest equivalent width not using the corresponding background quasars located in the vicinity of  foreground LRGs but instead random quasars with similar redshifts. In both cases we fix the width of the Gaussian line profile to be roughly the same as that of the actual measurement, in this case four pixels. The results of these null tests are shown in Figure~\ref{fig:nulltest}. Each panel shows the measurements for $12$ random realizations (gray diamonds).
In both cases the null test measurements are consistent with random noise and indicate that the detection of \mgii\ absorption shown in Table~\ref{tbl:correlation} and Figure~\ref{fig:correlation} is robust and not induced by systematic effects. These tests can also be used to estimate the intrinsic noise level of the statistical measurement.

To put our results in context, we first present existing measurements of the galaxy-metal absorption correlations for several species from the literature. This compilation is shown with open symbols in Figure~\ref{fig:correlation}. The magenta triangles are measurements for the \siiv$\,\lambda1393$ around Lyman-break galaxies (LBGs) at $z\sim2.2$ by \citet{steidel10a}, who also reported measurements for \lymana, \siii$\,\lambda1260$, \cii$\,\lambda1334$, \siii$\,\lambda1526$, \civ$\,\lambda1549$ and \alii$\,\lambda1670$ on similar scales (not shown to avoid crowdedness). \citet{bordoloi11a} measured the mean \mgii\ absorption around different types of galaxies at $z\sim0.7$. The orange stars show their measurements around red massive galaxies with stellar mass $M_*>10^{10.7}\,\MSun$ (though still about $0.5\,$ dex less massive than the LRGs used in this study). The gray circles show the mean \caii\ absorption around all galaxies at $z\sim0.1$ measured by \citet[][]{zhu13b}. Note this compilation is inhomogeneous in terms of galaxy types and redshifts but it shows the range of scales accessible to previous studies. The present analysis extends the detectability of the galaxy-gas (metal) correlation function up to about $20\,\mpc$, i.e. by two orders of magnitude.

The mean absorption profile does not show any cut off scale. The spatial correlation roughly follows a power law form of $r_\mathrm{p}^{-1.5}$. Such a slope implies a roughly constant S/N across all scales as the decrease in the signal amplitude is compensated by an increase in the number of usable pairs. This property allows measurements of 2-point correlation functions to reach large scales, such as in galaxy clustering and galaxy-galaxy lensing analyses. Our measurement allows us to probe the gas distribution around galaxies below and above the virial radius simultaneously. In Section~\ref{sec:halomodel}, we will interpret these measurements in the context of the standard cold dark matter model.

\subsection{From equivalent width to column density}\label{sec:columndensity}

\begin{figure}[t]
\epsscale{1.2}
\plotone{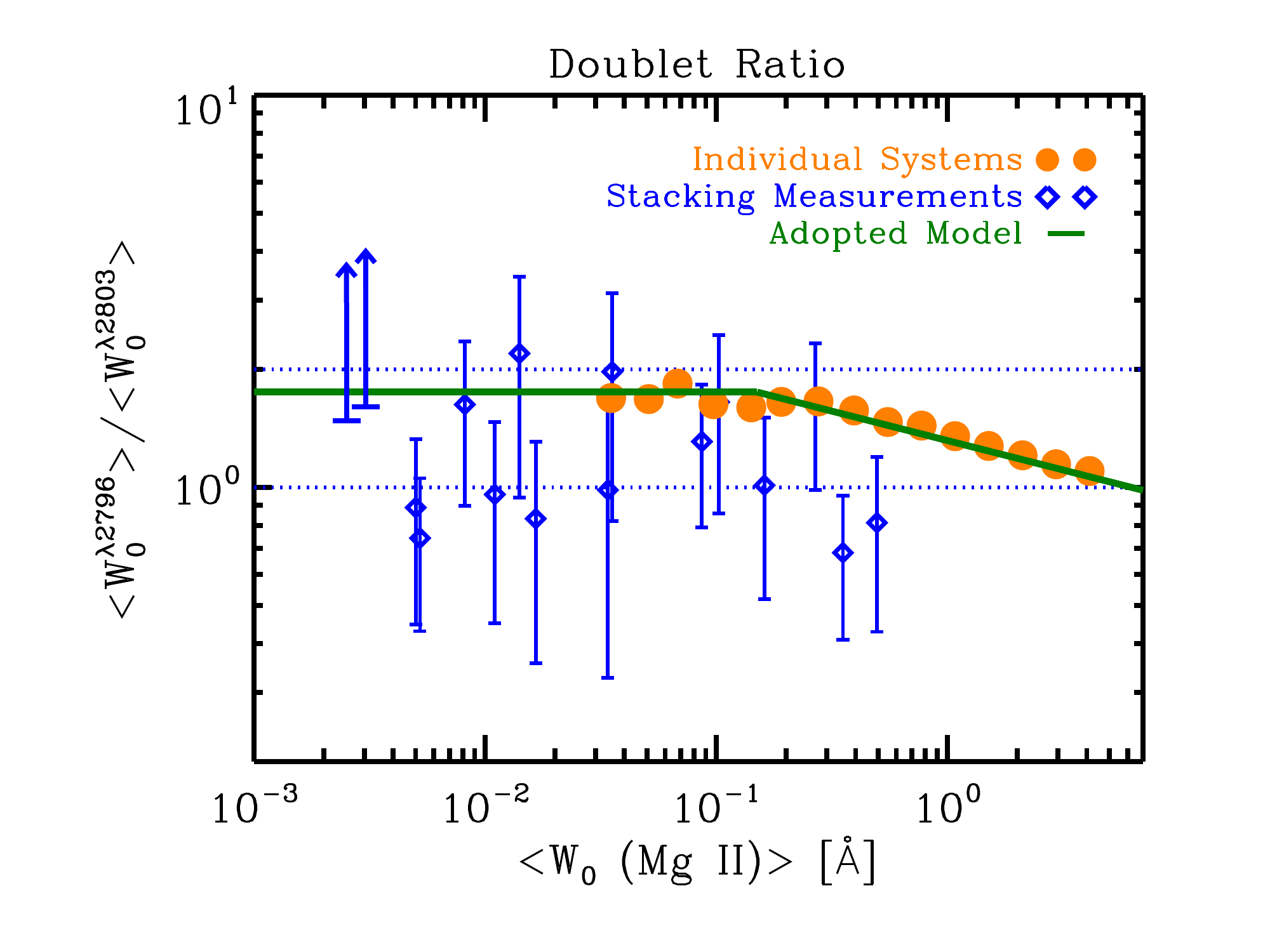}
\caption{
Doublet ratios as a function of \rewmgii. The orange points are median values of individual \mgii\ absorbers from \citet{churchill99a} and \citet{zhu13a}, and the green line is our adopted formula to capture the dependence on \rewmgii. The blue points are the measurements from the statistical analysis in this work. The two measurements on the far left are $2\sigma$ lower limits because the double-Gaussian fitting gives too small values of $\langle \rewmgiitwo \rangle$.
}
\label{fig:doubletratio}
\end{figure}

To estimate the surface density of magnesium from our mean measurements, we use the weaker of the two \mgii\ lines. From a measurement of the rest equivalent width of the full doublet, we estimate
\beq
\langle \rewmgiitwo \rangle = \frac{\langle \rewmgii \rangle}{1+DR} \,\mathrm{,}
\label{eq:weak_line_estimator}
\eeq
where $DR$ is the doublet ratio, bound between 1 and 2.

When absorption lines are not saturated we can directly infer gas column densities, as shown in Equation~(\ref{eq:curveofgrowth}). The saturation level depends on the column density and thermal broadening factor $b$. From high-resolution spectroscopic studies the thermal broadening factor of \mgii\ gas appears to be of the order of several $\kms$ \citep[\eg][]{churchill00a}. Taking $b$ to be $4\,$\kms, corresponding to $25,000\,$K, the stronger of the two \mgii\ lines starts to saturate when $\rewmgii \gtrsim 0.15\,$\AA. 

In the unsaturated regime, the \mgii\ surface density is given by (see Equation~\ref{eq:curveofgrowth}):
\beq
\langle \hat\Sigma_{\rm Mg\,II} \rangle = 
\frac{ 1.13 \times 10^{20}\,m_{\rm Mg}}{f_{2803}\,\lambda^2}\, \langle\rewmgiitwo\rangle~{\rm cm}^{-2} \,\mathrm{,}
\label{eq:sigma}
\eeq
where $m_{\rm Mg}$ is the atomic mass of magnesium.

\begin{figure*}
\epsscale{0.8}
\plotone{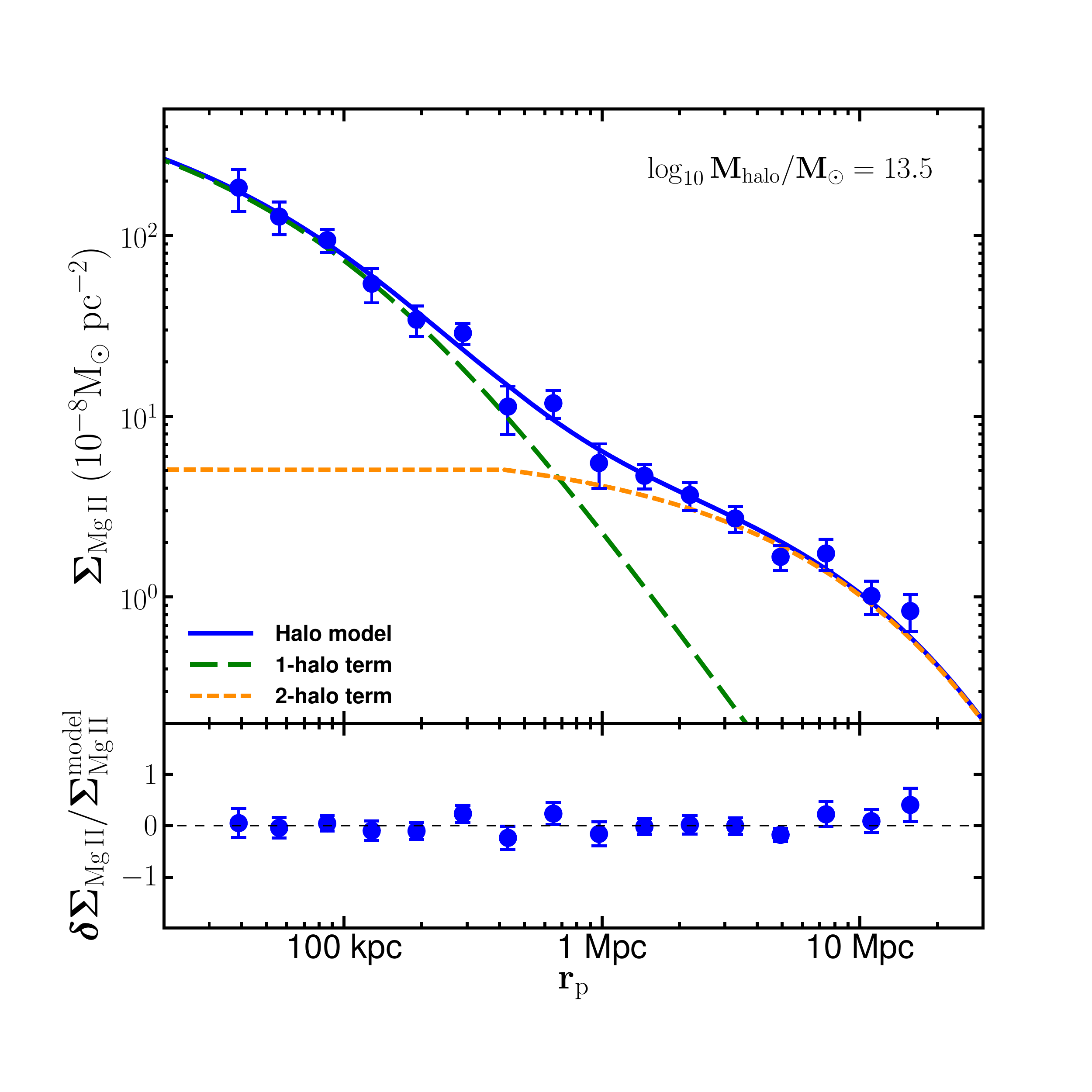}
\caption{The best-fit halo model. \emph{Upper} panel shows the best-fit halo model, decomposed into 1-halo and 2-halo terms. \emph{Lower} panel shows the fractional residuals. The halo model has three parameters: average LRG host halo mass $M_{\rm halo}$, \mgii\ gas-to-mass ratio in the host halo $f_{\rm Mg\,II}^{\rm 1h}$, and mean \mgii\ gas-to-mass ratio of all galaxies $f_{\rm Mg\,II}^{\rm 2h}$. 
}
\label{fig:halomodel}
\end{figure*}

On scales greater than about 200 kpc, our measurements show that $\langle \rewmgii \rangle < 0.1\,$\AA. In addition, our estimators show that the mean and median values are similar. This indicates that the fraction of saturated systems contributing to the overall signal is neglegible. In this regime we therefore expect a line ratio close to two.
This is in rough agreement with line ratio estimates of our stacked residual spectra, as shown in Figure~\ref{fig:doubletratio}. We note that the estimation of the line ratio of weak lines, detected a few orders-of-magnitude below the noise level of individual spectra, is difficult and possibly subject to systematic effects. Such line ratio estimates involve  measuring changes in the second-order moment of the (weak) stacked line profiles, as opposed to the rest equivalent width estimation which is based on the zero-th order moment of the line profile. It is therefore not surprising that the scatter of the measured line ratios is large.

Some authors have reported that in some cases weak absorbers with $\rewmgii<0.15\,$\AA\ can have line ratios smaller than $2$, indicating the strong line can still be saturated \citep[\eg][]{churchill00a}. We can obtain some guidance on the expected line ratio from direct detections of \mgii\ absorber systems. Using the individual absorber systems from \citet{churchill99a} and \citet{zhu13a}, we compute the median line ratio as a function of \rewmgii. This is shown with orange point in Figure~\ref{fig:doubletratio}. As expected we observe a break at around $\rewmgii \simeq 0.15 $\AA, below which the mean line ratio appears to be constant, with a value of about $1.75$. 
The similarity between the mean and the median ($\langle \rewmgii \rangle$) as a function of scale suggests that the fraction of saturated systems is scale independent. 
We therefore expect that the overall gas absorption is dominated by weak systems and $\langle \hat\Sigma_{\rm Mg\,II} \rangle \simeq\langle \Sigma_{\rm Mg\,II}^{\rm tot} \rangle$.

At $\rewmgii \gtrsim 0.15\,$\AA, a higher fraction of absorber systems is expected to occur.
As can be seen in Figure~\ref{fig:doubletratio} the median line ratio obtained from direct detections of absorbers reveals such a trend. To capture this behavior we adopt the following formula for absorbers in this regime:
\beq
\log_{10} DR = -0.15\;\log_{10} \frac{\langle \rewmgii \rangle}{0.15\,{\rm \AA}} + \log_{10} 1.75 \,\mathrm{,}
\label{eq:lineratio}
\eeq
which is shown with the green line in the figure.
In this regime, we estimate the \mgii\ surface density estimator using Equation~(\ref{eq:sigma}) with the line ratio provided by the above relation. On the corresponding scales, i.e. at $r_p<200\,$kpc, the fraction of saturated systems is expected to increase compared to that on larger scales. Our surface density estimate is therefore a \emph{minimum} value of the total surface density. 
In Appendix~\ref{sec:appendix:saturation}, we investigate the effect of different line ratio treatments 
and show that our conclusions are not strongly affected by this consideration.

\subsection{The velocity-space galaxy-gas correlation}\label{sec:vdisp}%

The galaxy-gas correlation function measured above is the projected surface density integrated along the line of sight, i.e., in the redshift (velocity) space. The velocity width of the absorption lines measured in the statistical analysis provides dynamical information of gas clouds around galaxies. The mean absorption line includes contributions from a large number of clouds and its width reflects the velocity dispersion of these clouds. 

We present the velocity dispersion measurements in Figure~\ref{fig:vdisp}. The velocity dispersion of \mgii\ gas clouds increases from about $100\,\kms$ at $30\,\kpc$ to about $700\,\kms$ at $20\,\mpc$. This is consistent with theoretical expectations. On small scales, the gas clouds are mostly from the LRG host halos and the velocity dispersion reflects their motion within the halo, while on larger scales, the gas clouds reside in neighboring dark matter halos and the velocity dispersion is determined by the motion of the neighboring halos, including the Hubble flow due to the expansion of the universe. We will discuss the measurements in more detail in the CDM cosmological context in Section~\ref{sec:halomodel:vdisp}.


\section{Interpretation}\label{sec:halomodel}

\subsection{The galaxy-gas correlation with the halo model}\label{sec:halomodel:correlation}

We now model the observed galaxy-gas correlation function. The measurement presented in Figure~\ref{fig:correlation} shows the mean \mgii\ rest equivalent width as a function of impact parameter, ranging from about 30 kpc, where most of the gas is expected to lie within the host dark matter halo of the LRGs, to several megaparsecs where most of the gas is expected to be associated with galaxies in neighboring halos. To describe the gas distribution over the entire range of scales, we make use of the dark matter halo model, originally developed to model the galaxy-mass and galaxy-galaxy correlation functions \citep[for a review, see][]{cooray02a}.

The dark matter halo model assumes that halo properties, such as density profile, abundance and galaxy occupation are determined solely by the halo mass. Here we extend this assumption to the gas distribution: we consider the gas-to-mass ratio $f_{\rm gas}$ to depend only on halo mass. This implies that, on average, the gas density profile in a halo with virial mass $M$ has the same NFW shape as dark matter up to an overall normalization determined by $f_{\rm gas} (M)$. The halo model we use has three parameters:
\begin{itemize}
\item the average virial mass $M_{\rm halo}$,
\item the gas-to-mass ratio $f^{\rm 1h}_{\rm gas} (M_{\rm halo})$ of the host dark matter halos (the 1-halo term),
\item the mean gas-to-mass ratio $f^{\rm 2h}_{\rm gas}$ in the CGM of all galaxies at $z\sim0.5$ (the 2-halo term). 
\end{itemize}
In this framework the mean gas surface density around galaxies is given by
\beq
\Sigma_{\rm gas} (r_\mathrm{p}) = f^{\rm 1h}_{\rm gas}(M_{\rm halo}) \Sigma^{\rm 1h}_{\rm m} (r_\mathrm{p}|M_{\rm halo}) +  f^{\rm 2h}_{\rm gas} \Sigma^{\rm 2h}_{\rm m} (r_\mathrm{p}|M_{\rm halo}) \, \mathrm{,}
\label{eq:halomodel}
\eeq
where the 1-halo term of the total surface density $\Sigma^{\rm 1h}_{\rm m} (r_\mathrm{p}|M_{\rm halo})$ is obtained by integrating the 3D NFW density profile along the line of sight and the 2-halo term $\Sigma^{\rm 2h}_{\rm m} (r_\mathrm{p}|M_{\rm halo})$ is calculated through the halo-mass cross correlation. Note that for simplicity we have dropped the ensemble average symbol. Galaxies can be central or satellite systems within a dark matter halo. LRGs being the most massive galaxies in the universe, we further assume all of them are central galaxies and the average mass of their host halos is $M_{\rm halo}$. We have tested that if a small fraction ($\sim10\%$) of LRGs are satellite systems, our conclusions on galaxy-gas and galaxy-mass correlations below are not affected, unless the gas-to-mass ratio of the host halos of these satellite LRGs are orders-of-magnitude higher than other halos. We present a detailed prescription of our halo model in Appendix~\ref{sec:appendix:halomodel}.

\begin{figure}[t]
\hspace{-1.55cm}
\epsscale{1.3}
\plotone{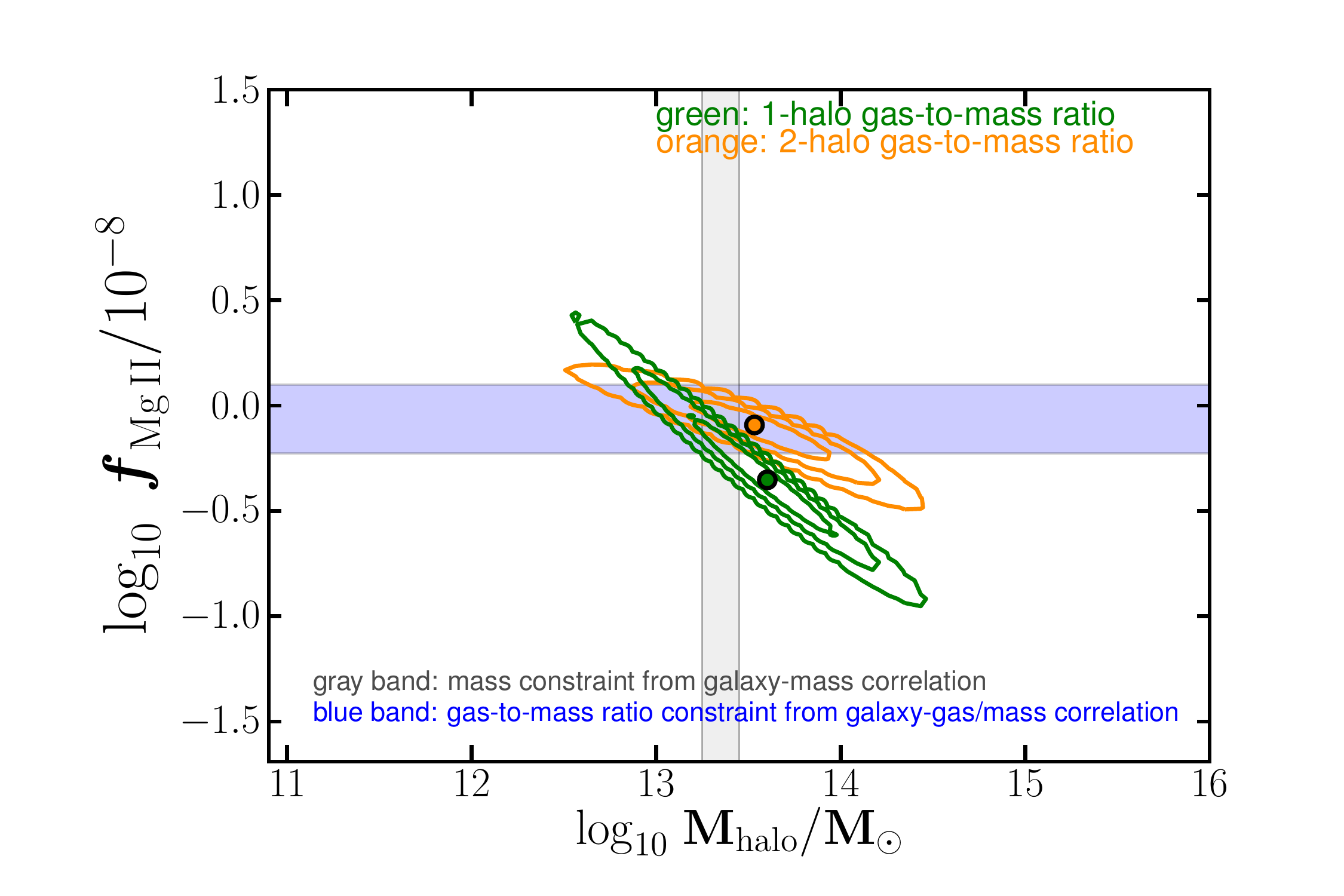}
\caption{Joint likelihood distributions for halo mass and gas-to-mass ratios. The contours indicate $1\sigma$ ($68.3\%$), $2\sigma$ ($95.4\%$), and $3\sigma$ ($99.7\%$) confidence intervals.}
\label{fig:chi2}
\vspace{.4cm}
\end{figure}

\begin{figure}[t]
\hspace{-1.5cm}
\epsscale{1.4}
\plotone{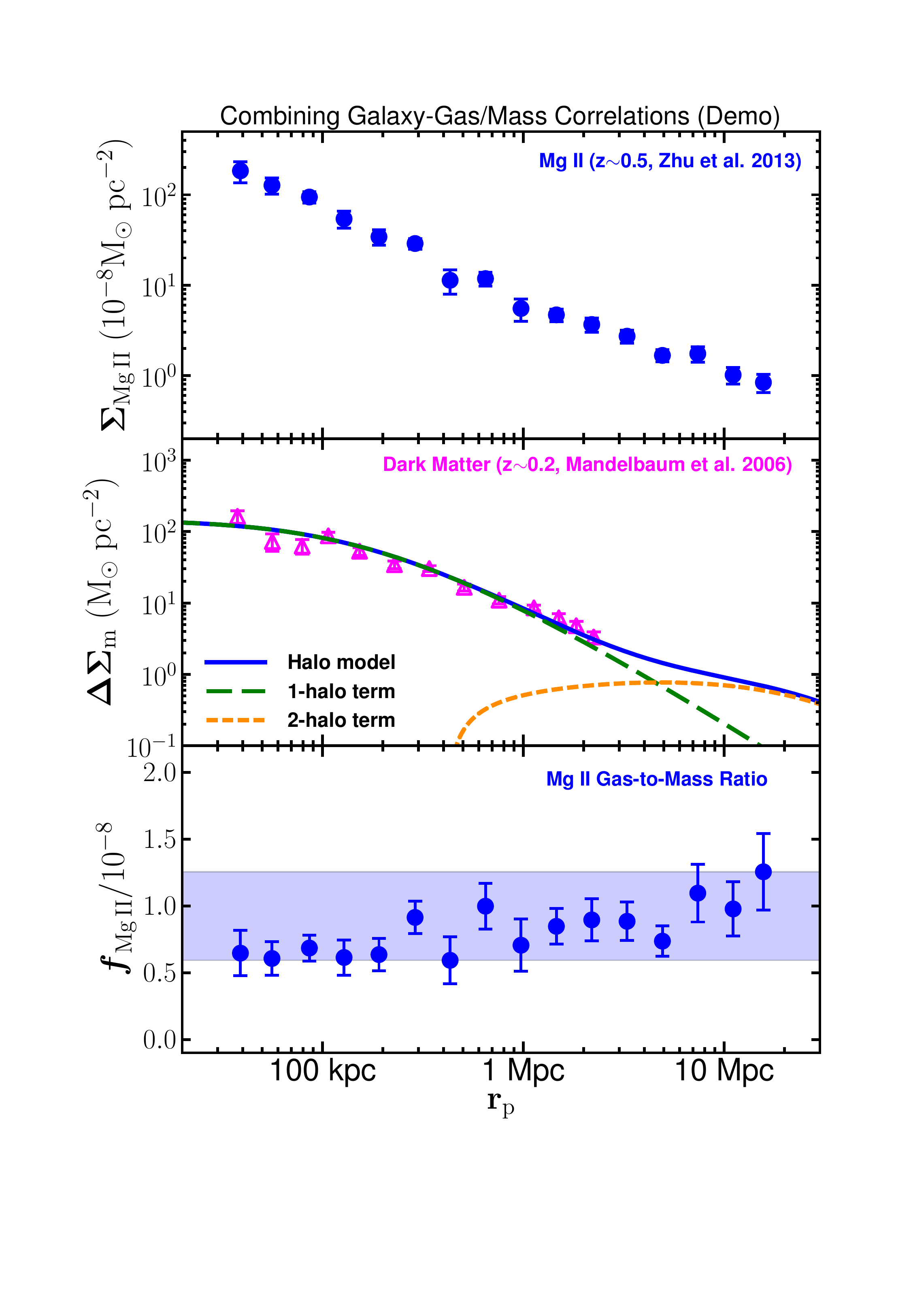}
\vspace{-2.0cm}
\caption{Direct constraints on the \mgii\ gas-to-mass ratio: the top panel shows the LRG-\mgii\ correlation function at $z\sim0.5$, the middle one shows the LRG-mass correlation by \citet[][]{mandelbaum06a} at $z\sim0.2$ from galaxy-galaxy lensing, with the lines showing the halo model with their best-fit halo mass $10^{13.5}\,\MSun$. The lower panel shows the ratio between these two quantities and provides us with a measurement of the \mgii\ gas-to-mass ratio as a function of impact parameter.
}
\label{fig:dark}
\vspace{.4cm}
\end{figure}

The halo model describes the mean projected surface density. As described in Section~\ref{sec:columndensity}, we adopt  $1.75$ for the line ratio when $\langle \rewmgii \rangle<0.15\,$\AA\ and Equation~(\ref{eq:lineratio}) otherwise, as suggested by individual systems. We then estimate the weaker line ($\lambda2803$) strength and the \mgii\ column density applying the linear relation of the curve of growth, Equation~(\ref{eq:curveofgrowth}).

\begin{figure*}[t]
\epsscale{1.1}
\plotone{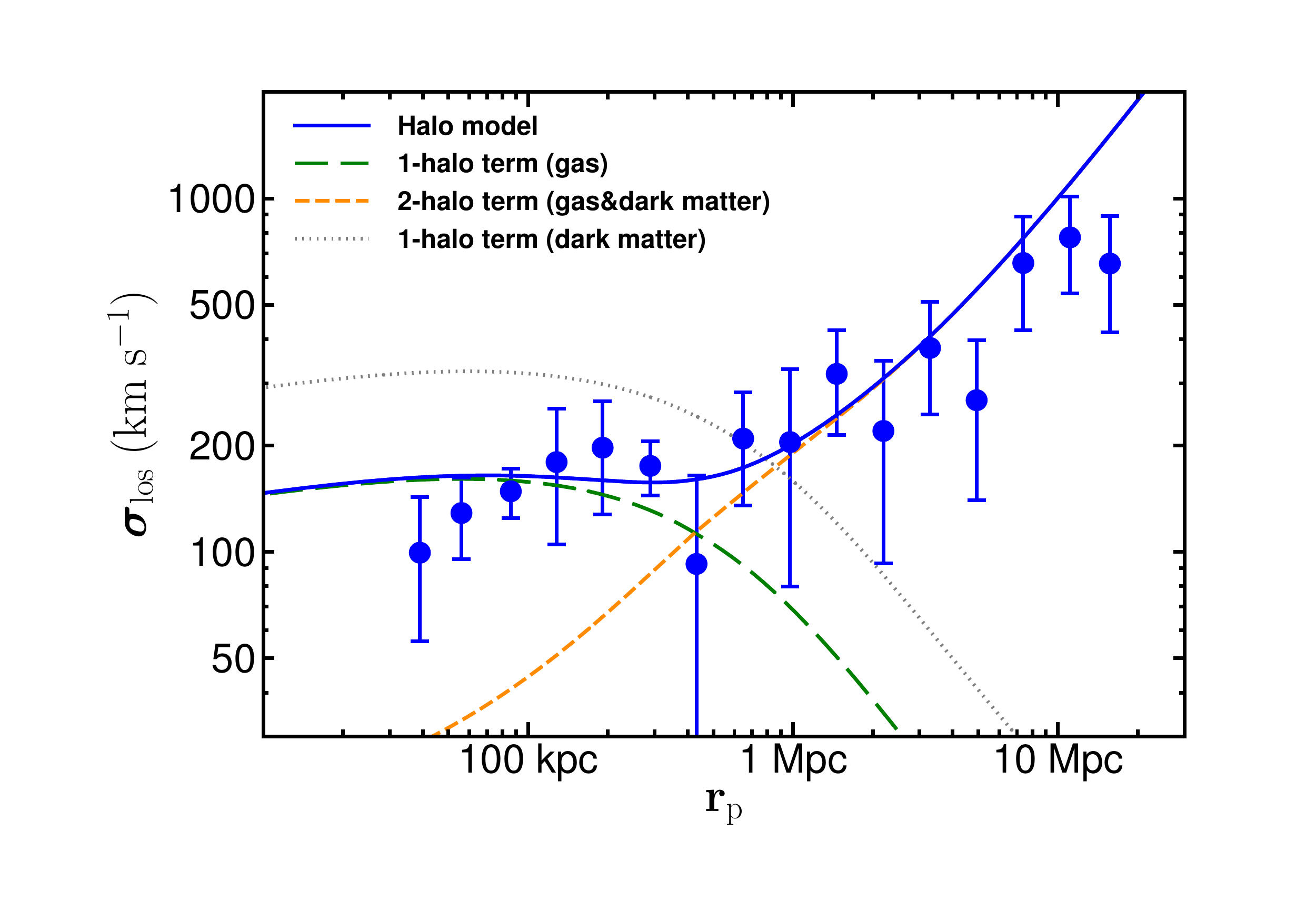}
\caption{The velocity dispersion of gas clouds traced by \mgii\ absorption. The lines are the halo model decomposed into 1-halo and 2-halo terms. With the halo mass ($10^{13.5}\,\MSun$) fixed, there is only one free parameter in the model, the velocity bias $\mu \equiv \sigma_{\rm gas}/\sigma_{\rm m} \approx 1/2$.
}
\label{fig:vdisp}
\end{figure*}

We generate Monte Carlo simulations spanning the 3-parameter ($M_{\rm halo}$, $f^{\rm 1h}_{\rm Mg\,II}$, $f^{\rm 2h}_{\rm Mg\,II}$) space and find the best-fit model to be
\begin{align}
\log_{10} M_{\rm halo}/\MSun&=13.5^{+0.3}_{-0.3} \\
\log_{10} f^{\rm 1h}_{\rm Mg\,II}&=-8.3^{+0.2}_{-0.2} \\
\log_{10} f^{\rm 2h}_{\rm Mg\,II}&=-8.1^{+0.1}_{-0.1} \, \mathrm{.}
\end{align}
The reduced chi-square is $\chi^2/dof=$ $0.72$. The errors reflect $1\sigma$ confidence level and do not include uncertainties in the conversion from rest equivalent width to column density, which we present separately in Appendix~\ref{sec:appendix:saturation}. Figure~\ref{fig:halomodel} shows the best-fit halo model and the fractional residuals. The small residuals show how well this halo model with only three parameters fits the data across about three orders of magnitude in scale. In Figure~\ref{fig:chi2}, we show the joint likelihood distributions in the $M_{\rm halo}-f^{\rm 1h}_{\rm Mg\,II}$ (green) and the $M_{\rm halo}-f^{\rm 2h}_{\rm Mg\,II}$ (orange) subspaces. The halo mass and gas-to-mass ratios are degenerate because they affect the overall amplitude in the same direction.

The best-fit halo mass is in excellent agreement with constraints from the halo modeling of the LRG-LRG auto-correlation by \citet[][]{white11a}, who estimated the mean halo mass of BOSS LRGs to be about $2-4\times10^{13}\,\MSun$. 
Galaxy-galaxy lensing analyses for the BOSS LRG sample are not yet available. We therefore choose to compare our results to 
the findings of \citet[][]{mandelbaum06a} who used a sample of LRGs at redshift $z\sim0.2$ (red subsample 6).
This sample has a similar average stellar mass and number density as the BOSS LRGs and the host halo mass is also consistent with that obtained from the galaxy-galaxy correlation by \citet[][]{white11a}. The best-fit halo mass of this sample is $(2.3\pm0.6)\times10^{13}\,\MSun$, shown with the vertical gray band in Figure~\ref{fig:chi2}. The excellent agreement between the constraints from different correlations shows our dark matter-gas halo model, with the assumption that gas shares the same density profile as dark matter, works well in describing the cool gas distribution in the cosmological context.

\subsection{Gas-to-mass ratio from observations}\label{sec:galaxy-mass/gas}

Having showed that the halo model applied to the gas distribution around LRGs provides us with a halo mass estimate consistent with what is found with other methods, we now focus on the value of the \mgii\ gas-to-mass ratio inferred from the same fit and assumption that on average the distribution of \mgii\ gas follows that of dark matter.

To constrain the gas distribution with respect to mass from observations in a model-independent way, we can simply divide the observed galaxy-gas correlation (the projected gas density profile) by the observed galaxy-mass correlation (the projected mass density profile). To do so we use again the red subsample 6 at redshift $z\sim0.2$ in \citet[][]{mandelbaum06a}. The observations are presented in the top and middle panels of Figure~\ref{fig:dark}. Galaxy-galaxy lensing analyses probe the difference between the average surface density within a radius and the surface density at that radius: $\Delta \Sigma_\mathrm{m} (r_\mathrm{p}) = \Sigma_\mathrm{m} (<r_\mathrm{p}) - \Sigma_\mathrm{m} (r_\mathrm{p})$. For a direct comparison, we estimate the average surface density within an aperture $\Sigma_{\rm m} (<r_\mathrm{p})$ using their best-fit halo model and subtract the observable $\Delta \Sigma_\mathrm{m} (r_\mathrm{p})$ from it. The halo model is overlaid in the middle panel. As done previously, we ignore the effect of satellite systems. Because the impact parameter binning is different, we use the best-fit halo model for interpolation to estimate the projected surface density at a given impact parameter $\Sigma_\mathrm{m} (r_\mathrm{p})$\footnote{We note that, using magnification instead of shear would allow us to directly infer 
the surface density $\Sigma_{\rm m}$ rather than $\Delta \Sigma_{\rm m}$, and interpolation with the halo model would not be necessary.}. 

We present the observed gas-to-mass ratio as a function of impact parameter in the bottom panel of Figure~\ref{fig:dark}, where we have overplotted a horizontal light blue band to encompass the maximum and minimum value. We find the mean \mgii-to-mass ratio to depend only weakly on scale. It varies by roughly a factor of two over three orders of magnitude in radius.
This range of \mgii-to-mass ratio is also shown in Figure~\ref{fig:chi2} as a blue region. We find it to be consistent with the constraints obtained from the halo model of the galaxy-gas correlation. Having applied the halo model to the gas distribution and found halo mass and gas-to-mass ratios in agreement with other methods suggests that it might be possible to apply the halo model to galaxy-gas correlations to constrain the mass of dark matter halos.

We can now put strong constraints on the \mgii\ gas-to-mass ratio around LRGs, with consistent results from the halo modeling of the galaxy-gas correlation function itself, and the combination of the observed galaxy-gas and galaxy-mass correlations. We can conclude that, in the LRG host halos at redshift $0.5$, (i) the \mgii\ gas-to-mass ratio is scale independent, i.e., the average \mgii\ gas density profile follows the same NFW shape as dark matter; and (ii) the \mgii\ gas-to-mass ratio is the same as the cosmic value, which is about $10^{-8}$.

We first examine the measured value of the \mgii\ gas-to-mass ratio. We emphasize that on large scales, the 2-halo term $f^{\rm 2h}_{\rm Mg\,II}$ is the mean value in the CGM of \emph{all} galaxies in the universe at redshift $0.5$. 
Neglecting the possible evolution of \mgii\ abundance from redshift $0.5$ to present day, the value of $10^{-8}$ indicates 
\begin{eqnarray}
\Omega_{\rm Mg\,II}^{\rm CGM} &=& \Omega_\mathrm{m} \times f^{\rm 2h}_{\rm Mg\,II}\nonumber\\ 
 &\approx& 3\times10^{-9} \,\mathrm{.}
\end{eqnarray}
Taking the solar abundance of Mg (${\rm [Mg/H]}=4\times10^{-5}$) as the maximum, this means \mgii\ probes at least $0.2\%$ of total baryons in the universe. If the \mgii\ abundance is only $0.1$ solar, as in High-Velocity Clouds/Magellanic Stream, then it traces about $2\%\times(0.1/[{\rm Mg\,II/H}])$ of total baryons. In the 1-halo regime, the \mgii-to-mass ratio in the LRG host halos is the same as the cosmic value. Assuming $0.1$ solar abundance again, the cool gas traced by \mgii\ in the LRG host halos would be about $10^{11}-10^{12}\,\MSun$, comparable to the stellar mass in LRGs while much more than their interstellar gas content \citep[about $10^9\,\MSun$, \eg][]{oosterloo10a, young11a}.

\subsection{The gas cloud velocity dispersion with the halo model}\label{sec:halomodel:vdisp}

The velocity width of the mean absorption lines provides dynamical information of gas clouds around galaxies. We present the line-of-sight velocity dispersion measurements in Figure~\ref{fig:vdisp}. The velocity dispersion of \mgii\ gas clouds increases from about $100\,\kms$ at $30\,\kpc$ to about $700\,\kms$ at $20\,\mpc$. 
We now use the halo model to investigate such spatial dependence of the gas cloud velocity dispersion. We fix the best-fit halo mass $M_{\rm halo}=10^{13.5}\,\MSun$ and gas-to-mass ratios and constrain the motion of gas w.r.t. the predicted motion of collisionless dark matter. 

As the surface density, the total line-of-sight velocity dispersion is decomposed into 1-halo and 2-halo terms:
\begin{eqnarray}
\sigma^2_{\rm los} (r_\mathrm{p}|M) &=& \mu^2(M)A^{\rm 1h}(r_\mathrm{p}|M)\, \sigma^2_{\rm 1h, los} (r_\mathrm{p}|M) \nonumber \\
  &+& A^{\rm 2h}(r_\mathrm{p}|M) \,\sigma^2_{\rm 2h,los} (r_\mathrm{p}|M) \, \mathrm{,}
\end{eqnarray}
where $A$ is the mass contribution of each term: 
\begin{eqnarray}
A^{\rm 1h}(r_\mathrm{p}|M) &=& \frac{\Sigma^{\rm 1h}(r_\mathrm{p}|M)}{\Sigma^{\rm 1h}(r_\mathrm{p}|M)+\Sigma^{\rm 2h}(r_\mathrm{p}|M)}\,\mathrm{,}\ \nonumber \\
A^{\rm 2h}(r_\mathrm{p}|M) &=& \frac{\Sigma^{\rm 2h}(r_\mathrm{p}|M)}{\Sigma^{\rm 1h}(r_\mathrm{p}|M)+\Sigma^{\rm 2h}(r_\mathrm{p}|M)}\,\mathrm{,}
\label{eq:mu}
\end{eqnarray}
and $\mu \equiv \sigma_{\rm gas}/\sigma_{\rm m}$ is the velocity bias between gas and dark matter. Note $M_{\rm halo}$ is simplified to be $M$ above.

On scales less than about one Mpc, the velocity dispersion is dominated by the motion of particles within the host halo. This 1-halo term is obtained by solving Jeans Equation of the NFW density profile for the virial motion of dark matter. On larger scales, the 2-halo term is the width of redshift-space correlation function, determined by the statistics of peculiar velocities, which describe the relative motion of neighboring halos and of particles within them, with respect to the background comoving frame (i.e. the Hubble flow). We estimate each contribution in the standard linear theory. A detailed prescription of the halo model for velocity dispersion is presented in Appendix~\ref{sec:appendix:vdisp}. 

The halo model is presented with the observations in Figure~\ref{fig:vdisp}. 
On large scales, the velocity dispersion is dominated by the Hubble flow and varies roughly linearly with scale. The observed width of the \mgii\ absorption is in good agreement with the expectation from the theory of dark matter fluctuations, indicating the contributing gas clouds reside in neighboring dark matter halos. 
On small scales, we observe that the velocity dispersion of the \mgii\ gas clouds is smaller than the virial velocity dispersion of collisionless dark matter. This implies that \mgii\ clouds found within the virial radius of LRGs are gravitationally bound and will not escape. It also shows that these clouds do not trace satellite galaxies within the halo. Their slower motion might be due to the fact that they are subject to the pressure of the hot gas filling the halo.

To fully characterize the radial distribution of gas velocity dispersion, we fit the observational results with Eq.~\ref{eq:mu}, using the best-fit halo mass derived above. We find the velocity bias $\mu$ to be about $1/2$, i.e., the gas cloud velocity dispersion around LRGs is a factor of two smaller than that of dark matter. Finally, we point out that, having previously estimated the mean halo mass of LRGs, a model having only one free parameter, the velocity bias $\mu$, is able to fit the data across three orders of magnitude in scale.

\section{Summary}\label{sec:summary}

The phase-space distribution of baryons and in particular metals encodes key information of galaxy formation processes. Absorption line spectroscopy is a powerful tool to probe gaseous matter but on large scales around galaxies, where densities are low, the direct detection of absorber systems is challenging. In this paper we use a statistical approach aimed at measuring absorption lines typically weaker than the noise level of individual background sources. We present a measurement of the mean \mgiidoublet\ absorption around luminous red galaxies, based on cross-correlating the positions of about one million red galaxies at $z\sim0.5$ and the flux decrements induced in the spectra of about $10^5$ background quasars from the Sloan Digital Sky Survey (see also P\'erez-R\`afols et al., in prep). We use quasar continuum estimates from \citet[][]{zhu13a} with calibration improvements to remove large-scale, sub-percent variations. Our ability to measure the absorption signal over a broad range of scales allows us to interpret the phase-space distribution of the gas in a cosmological context. Our results are summarized as follows:
\begin{itemize}
\item We measure the LRG-\mgii\ correlation function from $30\,\kpc$, where gas is associated with the LRG host halo, to about $20\,\mpc$, where it is dominated by contribution from neighboring halos. This galaxy-gas correlation reveals a change of slope on scales of about 1 Mpc, consistent with the expected transition from a dark matter halo dominated environment to a regime where clustering is dominated by halo-halo correlations. We use the observed rest equivalent width as a function of scale to obtain an estimate of the gas surface density, taking into account mild saturation effects on the smallest scales.
%
%
\item We find the standard halo model to provide an accurate description of the gas distribution over three orders of magnitude in scale under the simple assumption that the average distribution of \mgii\ gas follows that of dark matter up to a gas-to-mass ratio. Only 3 parameters are needed to describe the full range of measurements: the average host halo mass $M_{\rm halo}$, gas-to-mass ratio in the host halo $f_{\rm Mg\,II}^{\rm 1h}$, and mean gas-to-mass ratio in all neighboring halos $f_{\rm halo}^{\rm 2h}$. We find that a halo mass $M_{\rm halo}=10^{13.5}\,\MSun$ provides an excellent fit to the data. This LRG host halo mass is in good agreement with the constraints from the galaxy-galaxy and galaxy-mass correlation functions. Moreover, we find $f_{\rm halo}^{\rm 1h}$ to be consistent with $f_{\rm halo}^{\rm 2h}$.
\item Combining observations of the galaxy-mass and galaxy-gas correlation functions we obtain \emph{direct} constraints on the gas-to-mass ratio around LRGs $f_{\rm halo}^{\rm 1,2h}$ and find it to be roughly scale independent. This implies that (i) the average cool gas density profile around LRGs is consistent with the NFW profile, (ii) the density of \mgii\ clouds around LRGs is consitent with the cosmic value, estimated to be $\Omega_{\rm Mg\,II}^{\rm CGM} \approx 3\times10^{-9}$.
\item 
From line-width estimates, we show that the velocity dispersion of the gas clouds also shows the expected 1-halo and 2-halo behaviors. On large scales the gas distribution follows the Hubble flow, whereas on small scales we observe the velocity dispersion of \mgii\ gas clouds to be lower than that of collisionless dark matter particles within their host halo, by a factor of two. This indicates that \mgii\ gas clouds are gravitationally bound to their host halos and likely are slowly falling in unless a pressure gradient is large enough to stop them.
\end{itemize}
These results provide us with a new set of constraints on the large-scale distribution of gas. Extending the analysis to other species and different types of galaxies will help understanding the cosmic baryon cycle.

Large and homogeneous surveys of the sky have allowed us to probe the distribution of matter in low-density environments. From the Sloan Digital Sky Survey only, we now have measurements of the galaxy-galaxy correlation function from clustering analyses \citep[\eg][]{zehavi05a}, the galaxy-mass correlation from gravitational lensing \citep[\eg][]{mandelbaum06a}, the galaxy-dust correlation from reddening measurements \citep{menard10a} and, from this paper, the galaxy-gas correlation function obtained by measuring statistical absorption by metals. The velocity-space distribution of galaxies is also measured with various surveys \citep[\eg][]{jing98a, conroy07a}. Our work extends these measurements to one tracer of the gas distribution. These correlation functions are successfully interpreted in the standard CDM cosmological context and provide us with a more complete description of the matter distribution around galaxies in the phase space.

Our analysis demonstrates the power and potential of absorption line studies using the ever-growing data from large surveys.
The methods we developed in \citet{zhu13a, zhu13b} and the present paper are generic and readily applicable to any large dataset from future surveys eBOSS \citep[][]{comparat13a}, BigBOSS \citep[][]{schlegel11a}, and PFS \citep[][]{ellis12a}. These surveys will provide large samples of different types of galaxies at higher redshift where more species are accessible from the ground and a golden opportunity to improve our understanding of the gas distribution and the cosmic baryon cycle. This work also shows that a detection of the baryon acoustic oscillation feature with \mgii\ absorption is within reach.

\acknowledgments

This work was supported by NSF Grant AST-1109665, the Alfred P. Sloan foundation and a grant from Theodore Dunham, Jr., Grant of Fund for Astrophysical Research. We thank Jordi Miralda-Escud\'e for useful discussions and comments on the manuscript. We also thank Daniel Eisenstein for useful discussions.

Funding for the SDSS and SDSS-II has been provided by the Alfred P. Sloan Foundation, the Participating Institutions, the National Science Foundation, the U.S. Department of Energy, the National Aeronautics and Space Administration, the Japanese Monbukagakusho, the Max Planck Society, and the Higher Education Funding Council for England. The SDSS Web Site is http://www.sdss.org/. Funding for SDSS-III has been provided by the Alfred P. Sloan Foundation, the Participating Institutions, the National Science Foundation, and the U.S. Department of Energy Office of Science. The SDSS-III web site is http://www.sdss3.org/.

SDSS-III is managed by the Astrophysical Research Consortium for the Participating Institutions of the SDSS-III Collaboration including the University of Arizona, the Brazilian Participation Group, Brookhaven National Laboratory, University of Cambridge, Carnegie Mellon University, University of Florida, the French Participation Group, the German Participation Group, Harvard University, the Instituto de Astrofisica de Canarias, the Michigan State/Notre Dame/JINA Participation Group, Johns Hopkins University, Lawrence Berkeley National Laboratory, Max Planck Institute for Astrophysics, Max Planck Institute for Extraterrestrial Physics, New Mexico State University, New York University, Ohio State University, Pennsylvania State University, University of Portsmouth, Princeton University, the Spanish Participation Group, University of Tokyo, University of Utah, Vanderbilt University, University of Virginia, University of Washington, and Yale University.

\bibliographystyle{apj}


\begin{appendix}

\section{The dark matter-gas halo model}\label{sec:appendix:halomodel}

\subsection{The projected surface density}\label{sec:appendix:density}

The halo model \citep[\eg][and references therein]{ma00a, peakcock00a, seljak00a, scoccimarro01a, berlind02a, cooray02a} provides a simple tool linking observations of the large scale distribution of matter to theoretical predictions by dark matter cosmological models that is much less expensive than N-body/Hydrodynamic simulations. It was originally developed to investigate the galaxy and dark matter distribution, we here extend its use to the galaxy-gas correlation function.

We start with the formal definition of the projected galaxy-gas correlation function (Equation~\ref{eq:projectedomega}):
\beq
\omega_{\rm gal-gas} (r_{\rm p}) \equiv \left \langle \delta_{\rm gal} (r') \cdot \delta_{\rm gas} (r'+r_{\rm p}) \right \rangle \,\mathrm{,}
\eeq
where the ensemble average is performed over the entire area of interest. When the galaxy field is discretized, the ensemble average is restricted to the galaxy positions. The projected galaxy-gas correlation function is then equivalent to the \emph{excess} of the surface density, given by Equation~\ref{eq:surfacedensity}, which we rewrite here:
\beq
\langle \Sigma_{\rm gas} (r_{\rm p}) \rangle_{\rm gal} \equiv \overline{\Sigma}_{\rm gas}\,\omega_{\rm gal-gas} (r_{\rm p})\,\mathrm{.}
\eeq
Below we will drop the ensemble symbol for simplicity.

In the halo model, we divide the surface density into 1-halo and 2-halo terms:
\beq
\Sigma_{\rm gas} (r_\mathrm{p}) = \Sigma^{\rm 1h}_{\rm gas} (r_\mathrm{p}) + \Sigma^{\rm 2h}_{\rm gas} (r_\mathrm{p}
) \, \mathrm{.}
\eeq
For central galaxies, the 1-halo term is obtained by integrating the host halo density profile along the line of sight and the 2-halo term is computed through the cross-correlation function between the center position of the host halo and gas in other halos. For satellite galaxies, the 1-halo term includes contribution from its own host (sub-)halo and its parent halo, and the 2-halo term is again the contribution from neighboring halos. 
We assume all LRGs are central galaxies and will therefore only present the central-galaxy terms below. For an example of modeling the satellite contribution in the galaxy-mass correlation, we refer the reader to \citet{mandelbaum05a}.

The essential assumption of the dark matter halo model is that the properties (e.g., profile, density bias, abundance, galaxy occupation) of a dark matter halo are solely determined by its mass $M$ \citep[\eg][]{press74a, sheth99a}. Though it has been shown recently that the formation history also plays an important role \citep[the assembly bias, \eg][]{gao05a, zhu06a, wechsler06a}, we ignore this subtlety here. To apply the halo model to the galaxy-gas correlation function, we further assume that the gas-to-mass ratio ($f_{\rm gas}$) depends only on the halo mass, and does not depend on scale, i.e., the shape of the density profile is the same for gas and dark matter. The halo model we use has three parameters:
\begin{itemize}
\item the average virial mass $M$ of the host dark matter halos,
\item the gas-to-mass ratio $f^{\rm 1h}_{\rm gas} (M)$ of the host dark matter halos (the 1-halo term),
\item the mean gas-to-mass ratio $f^{\rm 2h}_{\rm gas}$ of all galaxies at the same redshift (the 2-halo term).
\end{itemize}
The mean excess of the gas surface density around galaxies then follows
\beq
\Sigma_{\rm gas} (r_\mathrm{p}|M) = f^{\rm 1h}_{\rm gas} (M)\, \Sigma^{\rm 1h}_{\rm m} (r_\mathrm{p}|M)+f^{\rm 2h}_{\rm gas}\, \Sigma^{\rm 2h}_{\rm m} (r_\mathrm{p}|M) \, \mathrm{,}
\label{eq:appendixhalomodel}
\eeq

We now present the ingredients for the 1-halo and 2-halo mass terms $\Sigma^{\rm 1h}_{\rm m} (r_\mathrm{p}|M)$ and $\Sigma^{\rm 2h}_{\rm m} (r_\mathrm{p}|M)$.

\noindent \textbf{$\bullet$ One-halo term:}
The 1-halo term is obtained by integrating the 3D density profile along the line of sight. We assume the dark matter density profile follows the NFW form \citep{navarro96a, navarro97a}:
\beq
\rho_\mathrm{m}(r|M)=\frac{\rho_\mathrm{s}}{(r/r_\mathrm{s})^{\gamma}(1+r/r_\mathrm{s})^{3-\gamma}}\,\mathrm{,}
\eeq
where $\gamma=1$. We express the scale radius $r_\mathrm{s}$ in terms of concentration $c$ and virial radius $\rvir$: $r_\mathrm{s}=\rvir/c$. The virial radius for a given halo mass $M$ is determined through
\begin{equation}
M=\frac{4\pi}{3}\bar{\rho}_\mathrm{m} \Delta_{\rm vir}\rvir^3\,\mathrm{,}
\end{equation}
where $\bar{\rho}_\mathrm{m}$ is the mean matter density and $\Delta_{\rm vir}$ is the critical overdensity for virialization, for which we adopt the fitting formula by \citet{bryan98a}:
\beq
\Delta_{\rm vir}(z) = \frac{1}{\Omega_\mathrm{m}(z)} \left\{ 18\pi^2+82\,[\Omega_\mathrm{m}(z)-1]-39\,[\Omega_\mathrm{m}(z)-1]^2 \right\} \,\mathrm{.}
\eeq
We assume the concentration $c$ follows
\begin{equation}
c(M,z) = \frac{c_0}{1+z} \left [ \frac{M}{M_{\star}} \right]^{-\beta} \, \mathrm{.}
\end{equation}
We take $c_0=9$ and $\beta=0.13$ \citep[\eg][]{bullock01a, hu03a}. The non-linear scale mass $M_{\star}=10^{12.7}\,\MSun$ for the adopted cosmology.
The scale density $\rho_\mathrm{s}$ is then determined through the integration of the profile:
\beq
M = \int_0^{\rvir} 4\pi r^2\, \rho_{\rm m}(r|M)\, \ud r  = \frac{4\pi\rho_{\rm s} \rvir^3}{c^3} \left[ \ln (1+c) - \frac{c}{1+c} \right ] \, \mathrm{,}
\eeq
where the second equal sign holds only for the NFW slope $\gamma=1$ (see \citealt{takada03a} for analytic formulae for other profiles).

To obtain the surface density, we integrate the NFW density profile along the line of sight:
\beq
\Sigma^{\rm 1h}_{\rm m} (r_\mathrm{p}|M) = \int_{-\infty}^{+\infty}\, \rho_{\rm m} \left(\sqrt{r^2_\mathrm{p}+s^2}|M\right)\, \ud s\,\mathrm{.}
\eeq
On large scales, the projected density profile follows $r_\mathrm{p}^{-2}$.

\noindent \textbf{$\bullet$ Two-halo term:} 
The 2-halo term is obtained by integrating the 3D cross-correlation function between the center position of the halo and matter of neighboring halos $\xi_{\rm hm}$:
\beq
\Sigma^{\rm 2h}_{\rm m} (r_\mathrm{p}|M) = \bar{\rho}_\mathrm{m} \int_{-\infty}^{+\infty} \xi_{\rm hm}(\sqrt{r^2_\mathrm{p}+s^2}|M)\, \ud s\, \mathrm{.}
\eeq
Note we have again dropped the background term so that this is the \emph{excess} of the surface density.
The correlation function $\xi_{\rm hm}$ in the halo model involves convolution of the halo-halo correlation function and the halo density profile. Since convolution in real space is simply multiplication in Fourier space, it is easier to calculate the power spectrum $P_{\rm hm}(k)$ first then obtain the correlation function by Fourier Transformation. The 2-halo power spectrum is given by
\beq
P_{\rm hm}(k) = b(M)P_{\rm lin}(k) \int_{M_{\rm min}}^{M_{\rm max}} \ud \nu f_\nu b(\nu) u(k|\nu)\,\mathrm{,}
\label{eq:phmk}
\eeq
where we have followed the convention and used the overdensity peak height $\nu$  \citep[\eg][]{bardeen86a}:
\beq
\nu \equiv \frac{\delta_c(z)}{D(z)\sigma(M)}\,\mathrm{.}
\eeq
Here $D(z)$ is the growth factor and $\delta_c(z)$ is the overdensity threshold for spherical collapse, for which we use the fitting formula given by \citet[][]{weinberg03a}:
\beq
\delta_{\rm c}(z) = \frac{3}{20} (12 \pi)^{2/3} \left[ 1 + 0.013 \log_{10} \Omega_{\rm m}(z) \right]\,\mathrm{.}
\eeq
The $\sigma(M)$ term is the present-day rms fluctuation in the mass density, smoothed with a top-hat filter of radius $R(M)\equiv(3M/4\pi\bar{\rho}_\mathrm{m})^{1/3}$:
\beq 
\sigma^2(M) = \int_0^{+\infty} \frac{\ud k}{k} \, \frac{k^3 P_{\rm lin}(k)}{2\pi^2} W^2 (kR) \, \mathrm{,}
\eeq
where $W$ is the Fourier transform of the top-hat window function:
\beq
W(x) = \frac{3}{x^3} (\sin x - x\cos x)\,\mathrm{.}
\label{eq:tophatwindow}
\eeq
For the large-scale bias $b$, we use the fitting formula:
\beq  
b(M, z) = b(\nu) = 1 + \frac{1}{\sqrt a \delta_{\rm c}} \left [ \sqrt a (a\nu^2) + \sqrt a b (a\nu^2)^{1-c} - \frac{(a\nu^2)^c}{(a\nu^2)^c +   b(1-c)(1-c/2)} \right] \, \mathrm{,}
\eeq
with $a=1/\sqrt{2}$, $b=0.35$, and $c =0.8$ \citep[][]{sheth01a, tinker05a}. The mass function $f(\nu)$ is defined as
\beq
\frac{\ud n}{\ud M} \ud M =  \frac{\bar{\rho}_{\rm m}}{M} f(\nu) \ud \nu\,\mathrm{,}
\eeq
and we use the fitting formula given by \citet{sheth99a}:
\beq
\nu f(\nu) = A \sqrt{\frac{2  a\nu^2}{\pi}} \left [ 1+ (a\nu^2)^{-p} \right] \exp \left ( -\frac{a\nu^2}{2} \right ) \, \mathrm{,}
\eeq
with $a = 0.707$ and $p = 0.3$. The coefficient $A$ is set by the normalization condition: 
\begin{equation}
\int_0^{+\infty} f(\nu) \ud \nu =1 \, \mathrm{,}
\end{equation}
and is $0.129$ in the cosmology we adopted. The $u(k|\nu)$ term is the Fourier transform of the density profile:
\beq
u(k|\nu) = \int 4\pi r^2 \ud r\, \rho(r|M) \frac{\sin kr}{kr}\,\mathrm{.}
\eeq
For the linear power spectrum $P_{\rm lin}(k)$, we use the fitting formula given by \citet{eisenstein99a}. Note the power spectrum is given in comoving space and after the Fourier transformation we convert the correlation function into physical space. 

The integral of the halo-mass power spectrum (Equation~\ref{eq:phmk}) is performed from $M_{\rm min} = 10^3\,\MSun$ to $M_{\rm max} = 10^{17}\,\MSun$. On large scales, the integral must equal one, so we also scale the integral such that it satisfies this condition. For metals, there is a lower halo mass limit below which no stars can form and metals can only come from stars formed in other halos \citep[][]{rees86a}. It is yet unknown what this lower limit is \citep[\eg][]{gnedin00a, okamoto08a}, but it only affects the overall amplitude of the integral, which we force to be one on large scales, so we keep $M_{\rm min} = 10^3\,\MSun$. The upper limit could be adjusted to take into account the halo exclusion effect, which only affects the small-scale power where the 1-halo term dominates, so we keep $M_{\rm max} = 10^{17}\,\MSun$.

\begin{figure}
\epsscale{0.5}
\plotone{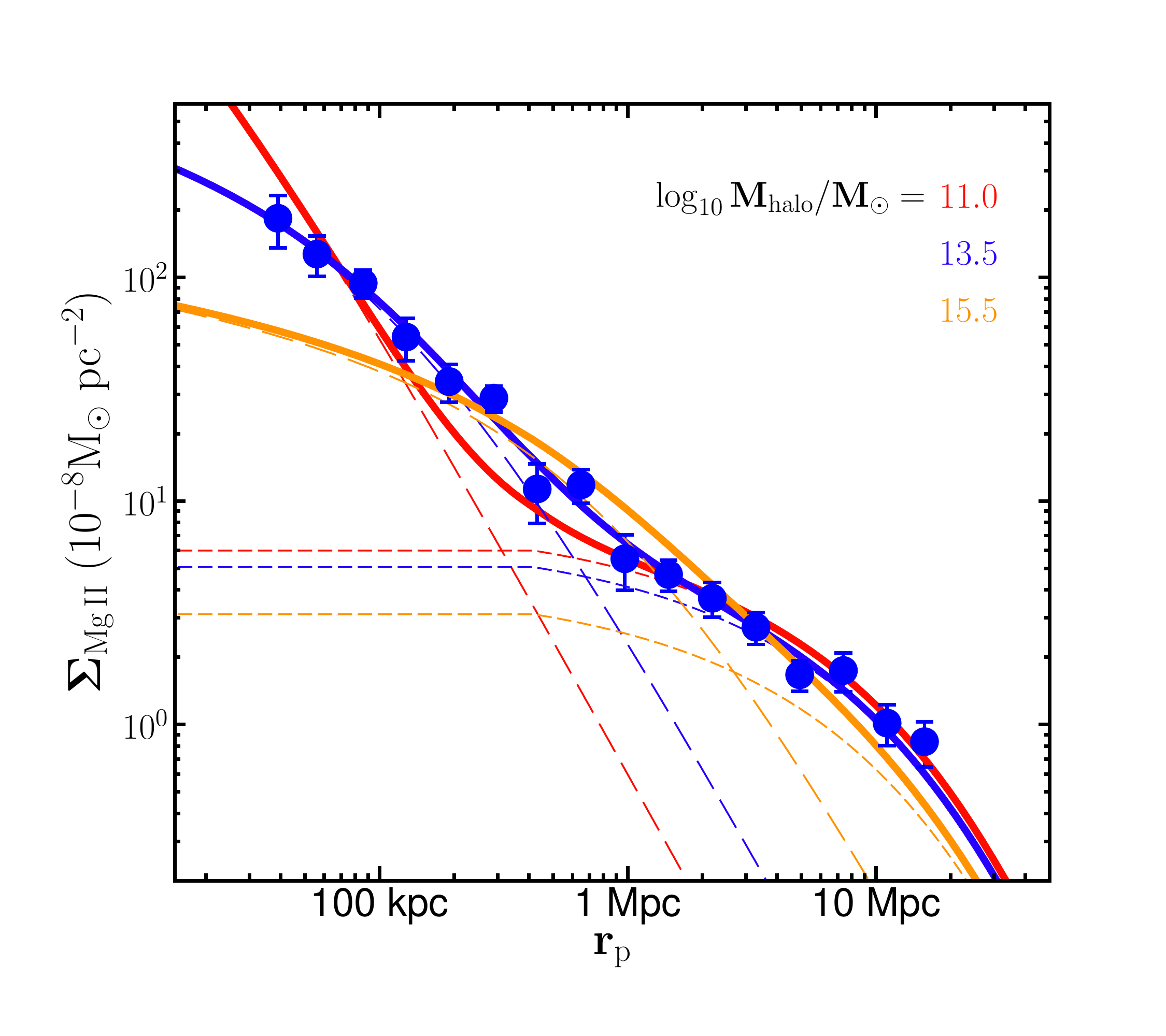}
\epsscale{0.5}
\plotone{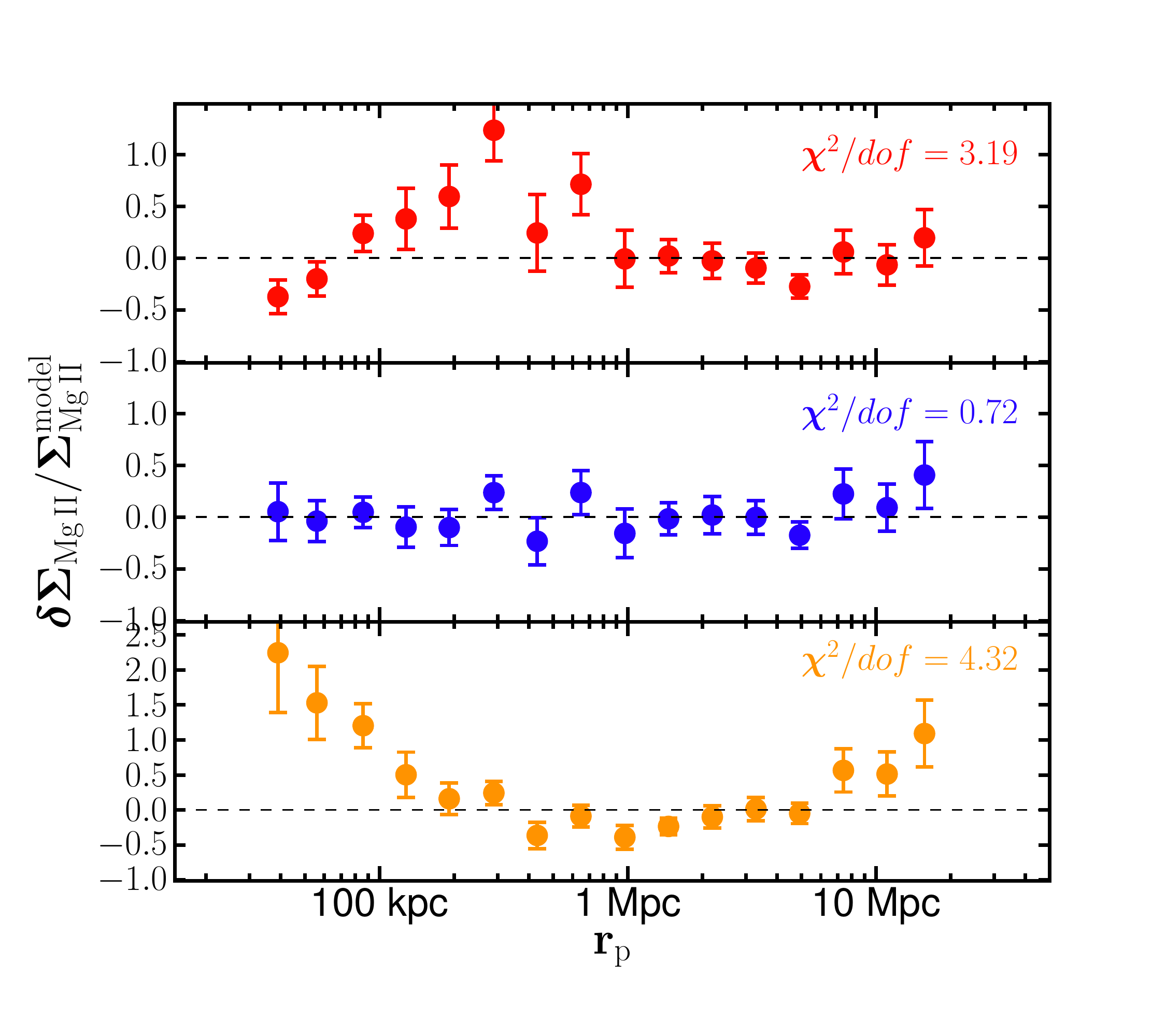}
\caption{The best-fit halo models (left panel) and fractional residuals (right panel) for three halo masses: $10^{11}\,\MSun$, $10^{13.5}$ (the best-fit mass), and $10^{15.5}\,\MSun$. The linestyles are the same as in Figure~\ref{fig:halomodel}, with different colors representing different masses.}
\label{fig:halodemo}
\end{figure}

Figure~\ref{fig:halodemo} shows examples of halo models with different halo masses, with the best-fit gas-to-mass ratios determined by minimizing the chi-square. For $M=10^{11}\,\MSun$, the lowest mass we probe, the profile is too steep on small scales and cannot capture the transition between the 1-halo and the 2-halo terms. For $M=10^{15.5}\,\MSun$, the largest mass we probe, the profile is too flat on small scales. The best-fit halo mass, $M=10^{13.5}\,\MSun$, provides an excellent fit to the measurements.

\subsection{The line-of-sight velocity dispersion}\label{sec:appendix:vdisp}

The total line-of-sight velocity dispersion of particles around a halo with mass $M$ is the mass-weighted summation, in quadrature, of the velocity dispersion of all particles along the line of sight:
\beq
\sigma^2_{\rm los} (r_\mathrm{p}|M)=\frac{1}{\Sigma(r_\mathrm{p}|M)}\int_{-\infty}^{+\infty}\, \rho_{\rm m} \left(\sqrt{r^2_\mathrm{p}+s^2}|M\right)\sigma^2_{\rm los} (r_\mathrm{p},s)\, \ud s \,\mathrm{,}
\eeq
where the line-of-sight dispersion at a given separation ($r_\mathrm{p}, s$) is given by combining the radial ($\sigma_\parallel$) and tangential ($\sigma_\perp$) dispersions:
\beq
\sigma^2_{\rm los} (r_\mathrm{p},s) = \sigma^2_\parallel \left(\sqrt{r^2_\mathrm{p}+s^2}\right) \sin^2 \theta + \frac{1}{2}\sigma^2_\perp \left(\sqrt{r^2_\mathrm{p}+s^2}\right) \cos^2 \theta \,\mathrm{,}
\eeq
with $\theta$ being the angle between the projection direction and the 3D separation, i.e., $\theta = \arctan (s/r_\mathrm{p})$. The radial ($\sigma_\parallel$) and tangential ($\sigma_\perp$) velocity dispersions are related by the velocity anisotropy: 
\beq
\beta = 1 - \frac{\sigma^2_\perp}{2\sigma^2_\parallel}\,\mathrm{.}
\label{eq:vdispanisotropy}
\eeq
The integral can be rewritten as
\beq
\sigma^2_{\rm los} (r_\mathrm{p}|M) = \frac{1}{\Sigma(r_\mathrm{p}|M)}\int_{-\infty}^{+\infty}\, \rho_{\rm m} \left(\sqrt{r^2_\mathrm{p}+s^2}|M\right) \left(1-\beta\frac{r^2_\mathrm{p}}{r^2_\mathrm{p}+s^2}\right)\sigma^2_\parallel \left(\sqrt{r^2_\mathrm{p}+s^2}\right)\, \ud s \,\mathrm{.}
\label{eq:losvdisp}
\eeq
We assume velocity isotropy, i.e., $\beta=0$, throughout this analysis.

We decompose the total line-of-sight velocity dispersion into 1-halo and 2-halo terms, as for the surface density:
\beq
\sigma^2_{\rm los} (r_\mathrm{p}|M) = \mu^2(M)A^{\rm 1h}(r_\mathrm{p}|M)\, \sigma^2_{\rm 1h, los} (r_\mathrm{p}|M) + A^{\rm 2h}(r_\mathrm{p}|M) \,\sigma^2_{\rm 2h,los} (r_\mathrm{p}|M) \, \mathrm{,}
\eeq
where $A$ is the mass contribution of each term: 
\beq
A^{\rm 1h}(r_\mathrm{p}|M) = \frac{\Sigma^{\rm 1h}(r_\mathrm{p}|M)}{\Sigma^{\rm 1h}(r_\mathrm{p}|M)+\Sigma^{\rm 2h}(r_\mathrm{p}|M)}\,\mathrm{,}\ 
A^{\rm 2h}(r_\mathrm{p}|M) = \frac{\Sigma^{\rm 2h}(r_\mathrm{p}|M)}{\Sigma^{\rm 1h}(r_\mathrm{p}|M)+\Sigma^{\rm 2h}(r_\mathrm{p}|M)}\,\mathrm{,}
\eeq
and $\mu \equiv \sigma_{\rm gas}/\sigma_{\rm m}$ is the velocity bias between gas and dark matter.

\noindent \textbf{$\bullet$ One-halo term:} 
The 1-halo term $\sigma^2_{\rm 1h}$ is obtained by solving Jeans Equation \citep[][]{binney87a}:
\beq
\frac{\ud \sigma^2_\parallel (r) \rho(r)}{\ud r} + \frac{2\beta(r)}{r}\sigma^2_\parallel (r) \rho(r) = -\rho(r) \frac{\ud \phi}{\ud r} = -\rho(r) \frac{GM(<r)}{r^2}\,\mathrm{.}
\label{eq:virialvdisp}
\eeq
For NFW profile and constant velocity anisotropy $\beta$, \citet{lokas01a} provides analytic solutions to the Jeans equation (Equation 13-16 in their paper). The velocity anisotropy has been shown to weakly depend on scale, increasing from around $0.15$ at small radius to about $0.4$ at virial radius \citep{colin00a, diemand04a}. The small anisotropy has little effect on the final line-of-sight velocity dispersion, so we assume velocity isotropy ($\beta=0$), in which case the radial velocity dispersion is given by Equation 14 in \citet{lokas01a}:
\begin{eqnarray}
\sigma^2_\parallel (x) &=& 
\frac{1}{2} V^2_{\rm vir}\,g(c) c x (1+x)^2 \times \\
&&
\left[\pi^2 - \log x - \frac{1}{x} - \frac{1}{(1+x)^2} - \frac{6}{1+x} + \left(1+\frac{1}{x^2}-\frac{4}{x}-\frac{2}{1+x}\right)\log(1+x) + 3\log^2(1+x)-6{\rm Li}_2(1+x)\right] \,\mathrm{,} \nonumber
\end{eqnarray}
where $x \equiv r/r_\mathrm{s}=cr/r_{\rm vir}$, $c$ is the concentration, $V_{\rm vir}$ is the circular velocity at virial radius:
\beq 
V_{\rm vir} = \frac{GM_{\rm vir}}{r_{\rm vir}}\,\mathrm{,}
\eeq
and
\beq
g(c) = \frac{1}{\log(1+c)-c/(1+c)}\,\mathrm{,}
\eeq
and ${\rm Li}_2$ is the dilogarithm: 
\beq
{\rm Li}_2 (z) = \int_1^z \frac{\log t}{1-t} \ud t\, \mathrm{.}
\eeq
The 1-halo term of the line-of-sight velocity dispersion can then be obtained by integrating Equation \ref{eq:losvdisp}. As we discussed in the main text, around LRGs this collisionless dark matter velocity dispersion is larger than the observed \emph{gas cloud} velocity dispersion by about a factor of two, i.e., $\mu_{\rm LRG}\approx1/2$.

\noindent \textbf{$\bullet$ Two-halo term:} 
The 2-halo term $\sigma^2_{\rm 2h}$ is the width of the correlation function in the redshift (velocity) space, and is determined by two factors: (1) the relative motion of the neighboring halos w.r.t. the host halo \citep{peebles80a, hamilton91a, mo97a, sheth01b, sheth01c} and of particles within these neighboring halos, w.r.t. the background comoving frame; (2) the Hubble flow with peculiar velocity, which determines the Kaiser limit of the redshift-space correlation function \citep[\eg][]{kaiser87a, hamilton92a}. We present these two terms separately below.   
Alternatively, one can also fold the relative motion of halos w.r.t. the background (the first term) into the Kaiser-limit redshift-space correlation function \citep[\eg][]{fisher95a, scoccimarro04a}. 

We first present the prescription of the first term. The velocity dispersion between the center of the host halo with mass $M$ and gas in another halo with mass $m$ at a distance $r$ (in 3D) can be decomposed into four terms:
\beq
\sigma^2_{Mm}(r) = \sigma^2_{\rm halo} (M) + \sigma^2_{\rm halo} (m) + \mu^2(m)\sigma^2_{\rm vir} (m) - 2\Psi_{Mm}(r)\,\mathrm{,} 
\eeq
where $\sigma_{\rm halo}(m)$ is the cosmic velocity dispersion of halos with mass $m$, $\sigma_{\rm vir}(m)$ is the mean virial motion of particles within the halo,
which can be obtained by solving the Jeans equation and taking the mass-weighted average, $\mu(m)$ is the velocity bias between gas and dark matter, and $\Psi_{Mm}(r)$ is the velocity correlation between two halos because their velocities are not independent. For the 2-halo term, we assume $\mu(m)$ to be one, but it has little effect since the virial motion of particles plays a sub-dominant role on scales where 2-halo term dominates. 

Following \citet{sheth01b}, the halo velocity dispersion from linear theory is given by
\beq
\sigma_{\rm halo} (m) = H_0f(\Omega_{\rm m})\sigma_{-1}\sqrt{1-\sigma^4_0/\sigma^2_1\sigma^2_{-1}}\,\mathrm{,}
\eeq
where $f(\Omega_{\rm m})=\ud \log D/\log a \approx \Omega_{\rm m}^{0.55}$ and
\beq
\sigma^2_{j}(m) = \frac{1}{2\pi^2} \int \ud k\, k^{2+2j} P(k) W^2[kR(m)]\,\mathrm{,}
\eeq
with $W(x)$ being the Fourier transform of the top-hat smoothing window, as Equation~\ref{eq:tophatwindow}. The square root term is to correct the fact that overdensities are not completely random patches.
For $\Omega_{\rm m}\sim0.3$, the halo velocity dispersion depends only weakly on mass and we use the fitting formula given by \citet{sheth01b}:
\beq
\sigma_{\rm halo} (m) = \frac{\sigma_{\rm fit}}{1+(R/R_{\rm fit})^\eta}\,\mathrm{.}
\eeq
For the adopted cosmology and at redshift $z=0.52$, we find $R_{\rm fit}=50\,\mpc$, and $\eta=0.85$, and $\sigma_{\rm fit}=400\,\kms$ provide a good fit.

The velocity correlation function from linear theory is given by \citet[][]{gorski88a} and \citet{sheth01c}: 
\beq
\Psi_{Mm,\parallel/\perp}(r) = \left[ H_0f(\Omega_{\rm m}) \right]^2\,(1-\sigma^4_0/\sigma^2_1\sigma^2_{-1})\,\frac{1}{2\pi^2} \int \ud k\, P(k) W[kR(M)]W[kR(m)]K_{\parallel/\perp}(kr)\,\mathrm{,}
\eeq
where for the radial ($\Psi_{Mm,\parallel}$) and tangential ($\Psi_{Mm,\perp}$) velocity correlations, 
\beq
K_\parallel(x) = \frac{\sin x}{x} - \frac{2}{x^3}\,(\sin x - x \cos x)\,\mathrm{,}\ 
K_\perp(x) = \frac{2}{x^3}\,(\sin x - x \cos x)\,\mathrm{,}
\eeq
respectively. The total velocity correlation is $\Psi_{Mm}(r)=\Psi_{Mm,\parallel}(r)+\Psi_{Mm,\perp}(r)$ and can be obtained by simply replacing $K_{\parallel/\perp}(x)$ with $K(x)=\sin x/x$.

To compute the total 3D velocity dispersion w.r.t to the host halo, we need to integrate $\sigma^2_{Mm}$ over all neighboring halos:
\beq
\sigma^2_{\rm 2h, 3D}(r|M) = \frac{\int\ud m\,n(m) m \left[1+\xi_{Mm}(r)\right] \sigma^2_{Mm}(r)}{\int\,\, \ud m\,n(m) m \left[1+\xi_{Mm}(r)\right]}\,\mathrm{.}
\label{eq:twohalo3Dvdisp}
\eeq
The line-of-sight velocity dispersion, ignoring the Hubble flow for the time being, is then obtained by inserting this quantity into the integral \ref{eq:losvdisp}:
\beq
\sigma^2_{\rm no\ Hubble} (r_\mathrm{p}|M) = \frac{1}{\Sigma^{\rm 2h}(r_\mathrm{p}|M)}\, \bar{\rho}_{\rm m} \int_{-\infty}^{+\infty}\, \xi_{\rm hm} \left(\sqrt{r^2_\mathrm{p}+s^2}|M\right) \frac{1}{3}\,\sigma^2_{\rm 2h, 3D}\left(\sqrt{r^2_\mathrm{p}+s^2}|M\right)\, \ud s \,\mathrm{,}
\label{eq:twohalolosvdisp}
\eeq
where we have assumed velocity isotropy ($\beta=0$). This equation involves a quadruple integral, one over $m$, one over $k$ for $\sigma^2_j$, another over $k$ for $\xi$, and one over $r$ along the line of sight at $r_\mathrm{p}$. In practice, we find that choosing a typical neighboring halo mass without doing the integral over $m$ can provide a good approximation, and tests show that the results are insensitive to the chosen halo mass between $10^{10}\,\MSun$ and $10^{14}\,\MSun$. This is because the halo velocity dispersion only weakly depends on mass, and the presence of the velocity correlation further cancels out most of the dependence. We therefore use a typical halo mass $10^{12}\,\MSun$ to circumvent the computational difficulty and do not perform the integral over $m$.

We now turn to the second term, the width of the Kaiser-limit redshift-space correlation function along the line of sight. We estimate this term by measuring the FWHM and dividing it by $2.35$, i.e., $\Delta v = {\rm FWHM}/2.35$. We model the correlation function with the standard spherical Legendre expansion method \citep[][]{kaiser87a, hamilton92a}, and empirically determine the width of its line-of-sight projection as a function of impact parameter. For the adopted cosmology, at redshift $z=0.5$, we find the following linear relation is a good approximation for the velocity width (FWHM/2.35):
\beq
\Delta v (r_\mathrm{p}) \approx 90\,\kms\,\frac{r_\mathrm{p}}{\mpc}\,+100\,\kms\,\mathrm{.}
\eeq
This approximation is valid between about $1\,\Mpc$ and $20\,\Mpc$ but over-estimates the width beyond $20\,\Mpc$. We do not go beyond $20\,\Mpc$ in this analysis.
 
The final 2-halo term of the velocity dispersion is then given by
\beq
\sigma^2_{\rm 2h, los}(r_\mathrm{p}|M) = [\Delta v (r_\mathrm{p})]^2 + \sigma^2_{\rm no\, Hubble} (r_\mathrm{p}|M) \,\mathrm{.}
\eeq

\section{Saturation Effects}\label{sec:appendix:saturation}

In Section~\ref{sec:halomodel}, we adopt line ratio $1.75$ when $\langle \rewmgii \rangle < 0.15\,$\AA, and Equation~(\ref{eq:lineratio}) otherwise, as suggested by the median line ratios of individual \mgii\ absorbers. We then convert the rest equivalent width of the weaker line ($\lambda2803$) to the \mgii\ surface density. 
We here investigate two different extreme line ratio treatments: (1) line ratio equals $1$ across all scales, i.e., the contributing absorption is all saturated; (2) line ratio equals $2$ at $\langle \rewmgii \rangle<0.15\,$\AA, and $1$ otherwise, i.e, the contributing absorption is all unsaturated at $\langle \rewmgii \rangle<0.15\,$\AA, and all saturated otherwise.

\begin{figure*}[!t]
\epsscale{0.54}
\plotone{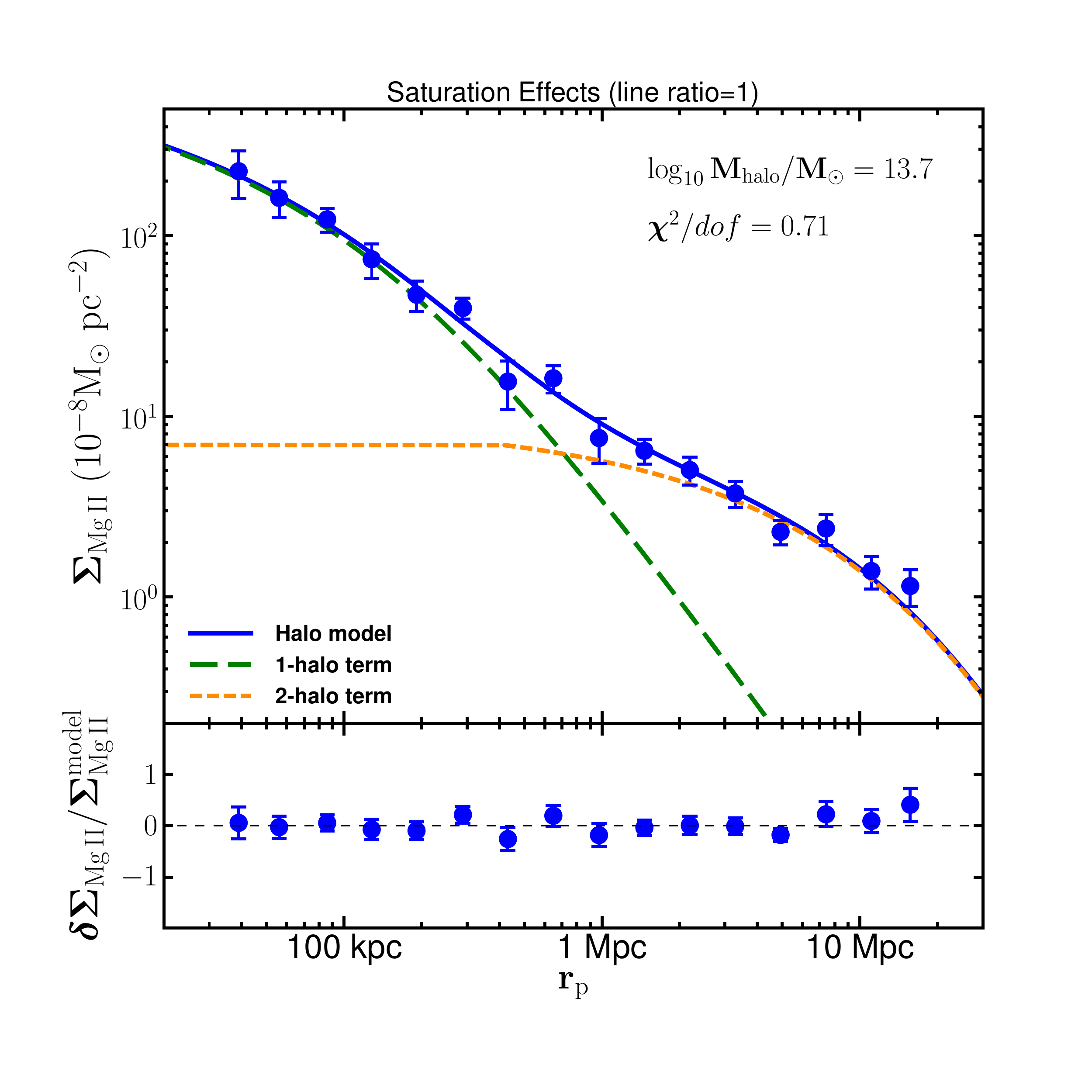}
\epsscale{0.54}
\plotone{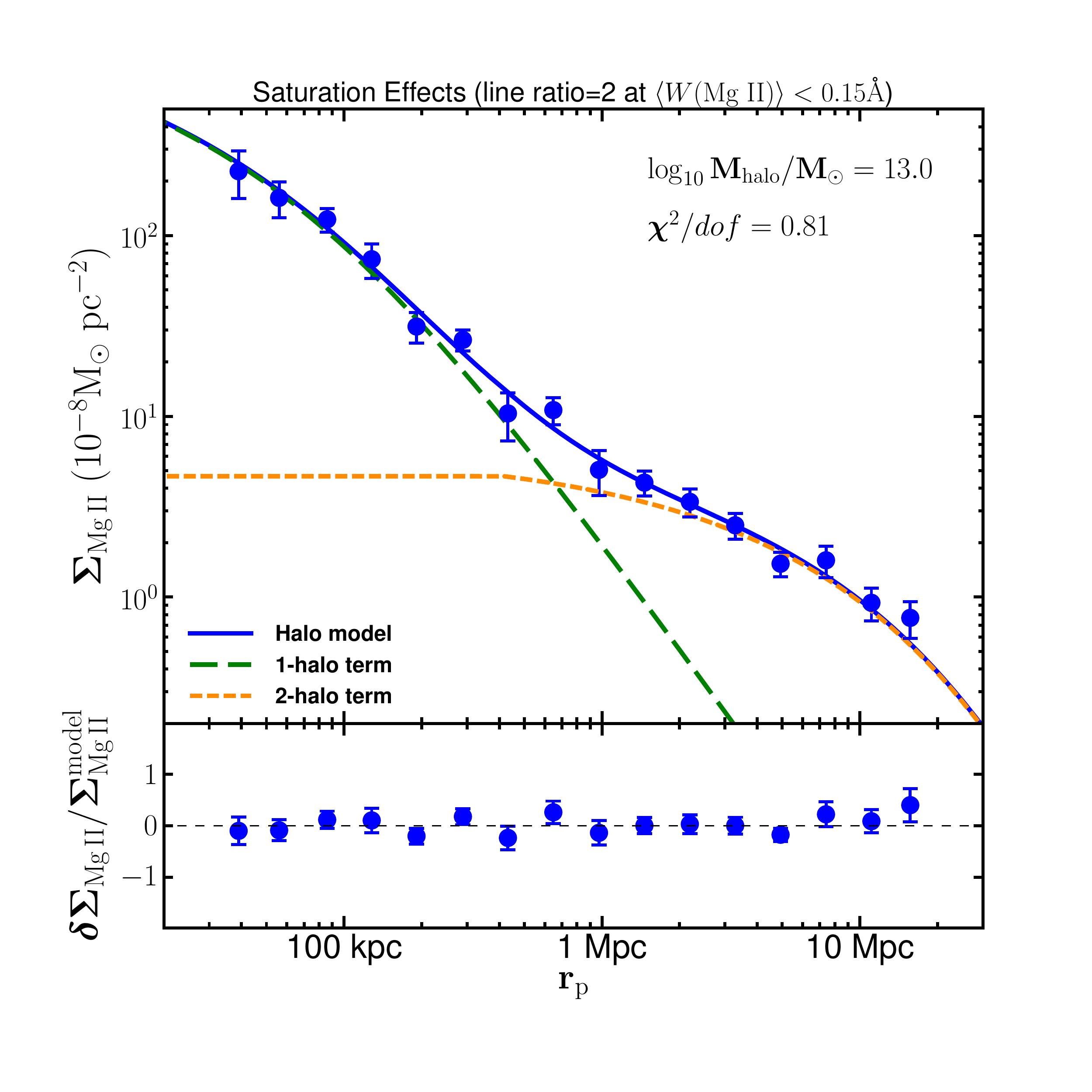}
\caption{Saturation effects on column density and halo modeling. The left panel shows the best-fit halo model if we adopt line ratio $1$ everywhere. The right panel shows the best-fit halo model if we adopt line ratio $2$ when $\langle \rewmgii \rangle < 0.15\,$\AA, and $1$ otherwise.
}
\label{fig:saturation-halomodel}
\end{figure*}

\begin{figure*}[!th]
\epsscale{0.56}
\plotone{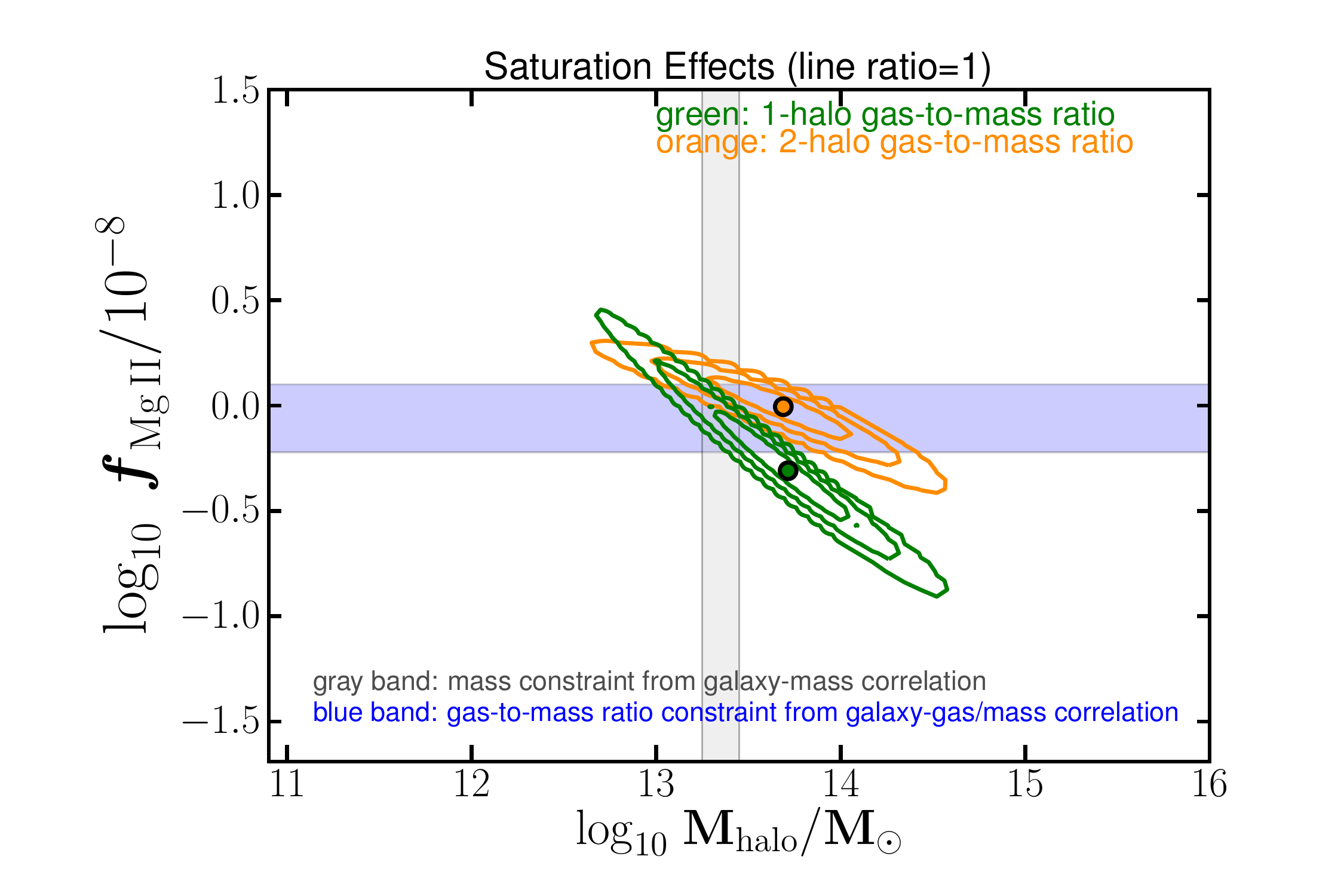}
\epsscale{0.56}
\plotone{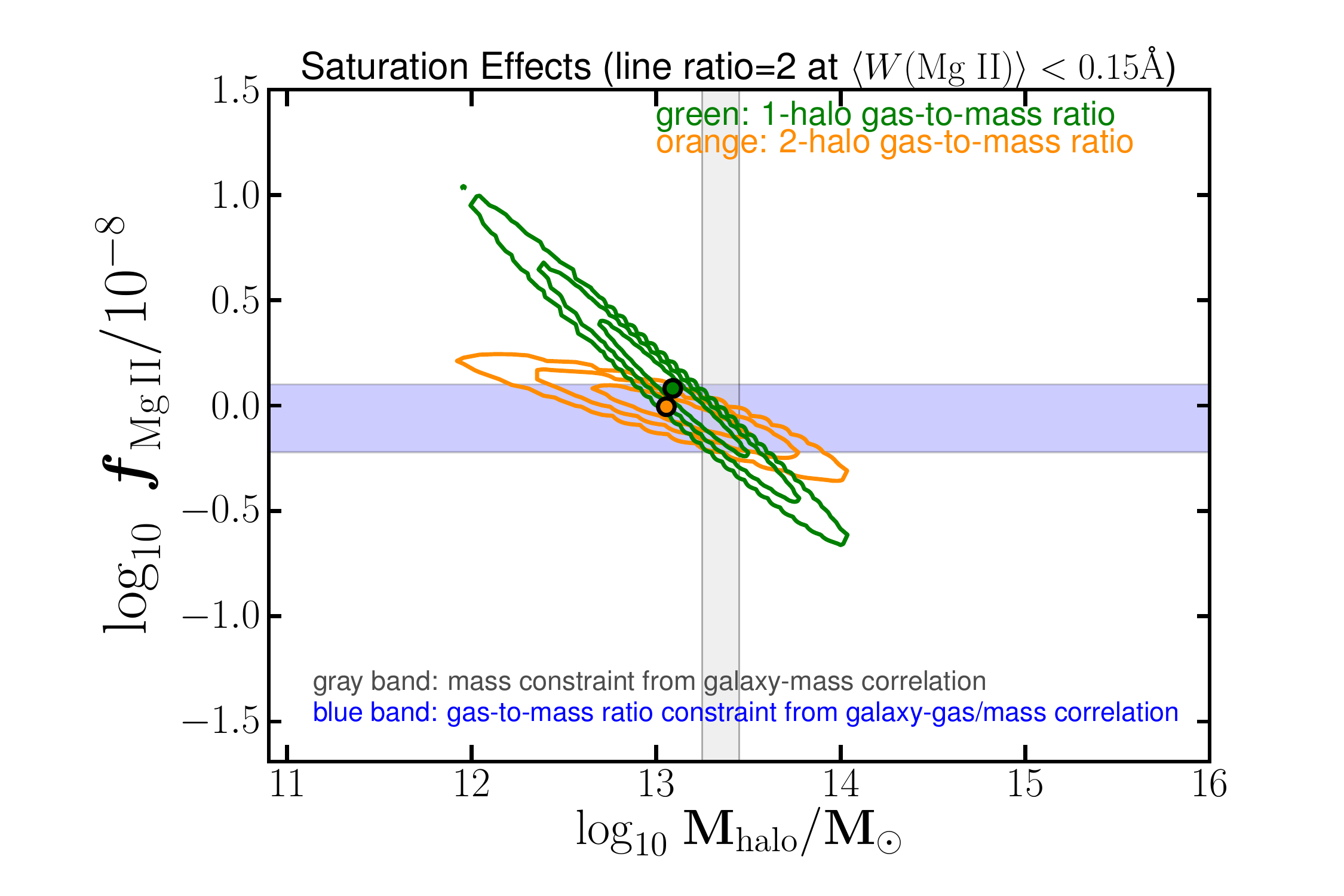}
\caption{Joint likelihood distributions of halo mass and gas-to-mass ratios for different line ratio treatments. The vertical gray band and horizontal blue band are the same as in Figure~\ref{fig:chi2}.
}
\label{fig:saturation-chi2}
\end{figure*}

The best-fit halo models with these two line ratio treatments are presented in Figure~\ref{fig:saturation-halomodel}. The joint likelihood distributions are shown in Figure~\ref{fig:saturation-chi2}. For comparison, we have also overplotted the same vertical gray and horizontal light blue bands as in Figure~\ref{fig:chi2}, the constraints from the galaxy-mass correlation. The best-fit halo parameters shift by about $0.2-0.5\,$dex ($1-2\sigma$), showing these extreme line ratio treaments do not have a significant effect on our conclusions.

It is worth pointing out that what we measure is the \emph{minimum} surface density because a fraction of the contributing absorbers must be saturated, even though accurate line ratio measurment and the choice of the weaker line can offer an estimate close to the true surface density. The saturation effect may become a major uncertainty when studying CGM of star-forming galaxies where we expect higher gas density \citep{bordoloi11a, tumlinson11a, zhu13b}. It is therefore necessary to develop more sophisticated models to include not only column density, but also Doppler broadening factor, covering fraction and other physical properties.

\end{appendix}

\end{document}